\documentclass[twocolumn,twocolappendix]{aastex701}
\usepackage{amsmath}
\usepackage{float}
\usepackage{soul}

\definecolor{blazeorange}{rgb}{1.0, 0.4, 0.0}
\definecolor{seagreen}{rgb}{0.18, 0.55, 0.34}
\definecolor{rufous}{rgb}{0.66, 0.11, 0.03}
\definecolor{royalfuchsia}{rgb}{0.79, 0.17, 0.57}
\definecolor{scarlet}{rgb}{1.0, 0.13, 0.0}
\definecolor{royalpurple}{rgb}{0.47, 0.32, 0.66}

\begin{document}

\title{Unexpectedly Weak General Relativistic Effects in Strongly Relativistic Tidal Disruption Events}

\author[0000-0003-2776-082X]{Ho-Sang Chan}
\altaffiliation{Croucher Scholar}
\affiliation{JILA, University of Colorado and National Institute of Standards and Technology, 440 UCB, Boulder, CO 80309-0440, USA}
\affiliation{Department of Astrophysical and Planetary Sciences, University of Colorado, 391 UCB, Boulder, CO 80309, USA}
\email[show]{hschanastrophy1997@gmail.com}

\author[0000-0003-2012-5217]{Taeho Ryu}
\affiliation{JILA, University of Colorado and National Institute of Standards and Technology, 440 UCB, Boulder, CO 80309-0440, USA}
\affiliation{Department of Astrophysical and Planetary Sciences, University of Colorado, 391 UCB, Boulder, CO 80309, USA}
\affiliation{Max-Planck-Institut für Astrophysik, Karl-Schwarzschild-Straße 1, 85748 Garching bei München, Germany}
\email{}

\author[0000-0002-2995-7717]{Julian Krolik}
\affiliation{Physics and Astronomy Department, Johns Hopkins University, Baltimore, MD 21218, USA}
\email{}

\author[0000-0002-7964-5420]{Tsvi Piran}
\affiliation{Racah Institute of Physics, The Hebrew University of Jerusalem, Jerusalem 91904, Israel}
\email{}


\begin{abstract}

Tidal disruption events (TDEs) occur when stars are destroyed by supermassive black holes and are among the brightest nuclear transients. It has been thought that strong relativistic effects rapidly dissipate orbital energy and produce prompt disk formation when the stellar pericenter is smaller than $\sim 10$ gravitational radii. Using a general relativistic hydrodynamic simulation of a strongly relativistic TDE involving a Sun-like star and a $10^{6}\,M_{\odot}$ non-spinning black hole, we find instead that the overall evolution is similar to weakly relativistic TDEs: the debris remains highly eccentric, with most of the returned mass residing near the orbital apocenter ($\sim 250\times$ the initial pericenter distance), and shocks, rather than accretion, power the event. The simulation starts from the initial stellar approach and follows the debris evolution up to $35$\,days after the peak mass-return time ($\simeq$ $23$\,days). Although early shocks driven by strong relativistic apsidal precession and pericenter nozzle compression dissipate orbital energy efficiently, they last only about a week ($\sim 0.3$ of the peak mass-return time). Stream self-interactions increase the incoming stream's angular momentum, thereby expanding its pericenter distance, weakening precession and shocks, and reducing dissipation. These results suggest that circularization in TDEs may proceed slowly regardless of the strength of apsidal precession, with the flow remaining highly eccentric and extended during the peak optical/UV luminosity.

\end{abstract}


\keywords{\uat{High energy astrophysics}{739} --- \uat{Black hole physics}{159} --- \uat{Hydrodynamical simulations}{767} --- \uat{General relativity}{641} --- \uat{Accretion}{14} --- \uat{Tidal disruption}{1696}}


\section{Introduction} \label{sec:intros} 

A tidal disruption event (TDE) occurs when a star approaches a supermassive black hole (SMBH) with a separation smaller than or comparable to the tidal radius $r_t \approx R_*(M_{\mathrm{BH}}/M_*)^{1/3}$, where $M_{\mathrm{BH}}$ is the SMBH mass, $M_*$ is the stellar mass, and $R_*$ is the stellar radius. At this distance, the tidal forces exerted by the SMBH become comparable to the star's self-gravity. For a zero-energy (parabolic) orbit, approximately half of the stellar debris remains gravitationally bound to the SMBH, while the other half escapes to infinity. The fallback rate of the bound debris rises to a peak value, $\dot{M}_0$, which can greatly exceed the Eddington mass accretion rate, and subsequently declines over time as $t^{-5/3}$ \citep{1988Natur.333..523R, 1989IAUS..136..543P}. This process produces a luminous flare far exceeding the luminosity of a quiescent black hole that is observable in the optical/UV and/or soft X‑ray bands \citep[see][for recent reviews]{2023arXiv231016879W,2025arXiv251114911M}.

The conventional view of TDE emission rests on the premise that the accretion rate onto the black hole tracks the mass fallback rate because the returning debris is assumed to rapidly `circularize', i.e., settle into circular orbits on the radial scale set by the debris' mean angular momentum, $\simeq 2\,r_p$. Accretion within such a disk leads to radiation, primarily in soft X-rays \citep{1988Natur.333..523R,2009MNRAS.400.2070S,2011MNRAS.410..359L}. Other works \citep{Balbus-Mummery2018,Mummery+2024} separate the accretion rate from the mass-return rate, while still asserting that the debris quickly `circularizes'.

However, this assumption is in strong tension with observations\footnote{Recent observations of the polarization in TDEs also support non-circular gas morphology and shock-powered emission \citep[e.g.,][]{Koljonen+2025,Floris+2026}.}: a large majority of TDE candidates are instead brightest at the optical/UV wavelengths \citep[e.g.,][]{2012EPJWC..3903001G, 2020SSRv..216..124V}. Moreover, the inferred sizes of the optical/UV emitting regions are typically $\sim 100 \times$ the expected dimensions of a circularized debris disk \citep{2021ARA&A..59...21G}. Furthermore, the orbital energy that must be dissipated in order to form a compact disk exceeds the observed radiated energy by several orders of magnitude \citep{2015ApJ...806..164P, 2017MNRAS.467.1426S}.

To resolve this tension, the hydrodynamics of the debris must be considered more carefully. The original argument for quick disk formation \citep{1988Natur.333..523R} held that strong relativistic apsidal precession would cause returning debris streams to intersect at a large angle, dissipating an amount of energy comparable to the debris' kinetic energy at $r \sim r_p$. However, later work \citep{2015ApJ...804...85S,Dai+2015,2020ApJ...904...68K} showed that for most TDEs, the pericenter is large enough that apsidal precession is relatively small, and the inter-stream shocks are therefore much weaker than initially envisaged. In global simulations since, for all of which the pericenter was $> 10\,r_g$, this expectation was upheld \citep{2023ApJ...957...12R,2024Natur.625..463S,2024ApJ...971L..46P, 2025arXiv250923894A}.

It is, however, of great interest to examine events with smaller pericenter distances to see what happens when the apsidal precession is strong. In such conditions, it has long been expected that the overall strength of shocks increases, and disk formation should be faster. 

To date, there have been only a handful of simulations of such events that start from realistic model parameters and track both the star’s disruption and the return of mass to the neighborhood of the black hole. \citet{2022MNRAS.510.1627A} studied the debris in the disruption of a $1$\,$M_{\odot}$ star by a $10^6$\,$M_{\odot}$ SMBH with a pericenter distance of $r_p \simeq 7$\,$r_g$, and followed it for $\lesssim 7$\,days, or $\lesssim 0.3$ of the debris' characteristic return time. In this short period of time, they found only limited circularization: the eccentricity of the debris remained high ($\simeq 0.88$) and was changing very slowly by the end of their simulation. More recently, \citet{2024ApJ...971L..46P} reported on simulations comparing non-relativistic to relativistic TDEs (i.e., pericenter distances $r_p = 50 \, r_g$ and $10\, r_g$)\footnote{The relativistic case reported in greatest detail assumed a Schwarzschild spacetime, but they also performed another one for spin parameter $0.99$ and a stellar orbit tilted with respect to the black hole spin axis.}. They found that disk formation in the relativistic case proceeds more rapidly than in their weakly relativistic simulations, but at the end of their simulation, roughly $10\times$ the time of peak mass-return, very nearly all the debris mass remained at radii $\gtrsim 10^3$\,$r_g$.

In this work, we investigate how the debris morphology in strongly relativistic TDEs evolves, from the star’s initial approach to the black hole to well past the peak mass-return time. Our general relativistic hydrodynamic (GRHD) simulation demonstrates that strong relativistic effects initially lead to the formation of strong shocks as expected, but this phase is short-lived, lasting only until $\sim 0.3$ of the peak mass-return time, which is about a week for our simulation's parameters. Thereafter, the shocks steadily weaken, and disk formation slows down. The debris orbits remain eccentric to well past the peak mass-return time, with most of the returned mass residing near the orbital apocenter. At this point, dissipation of orbital energy is so slow that the formation of the circular disk on the scale of $r_p$ would take at least an order of magnitude more time. 

This paper is organized as follows. \S~\ref{sec:methods} describes the simulation methodology, covering both the stellar disruption and debris fallback phases. \S~\ref{sec:results} presents the main findings. \S~\ref{sec:discuss} discusses the implications of our results, compares them with previous studies, and summarizes our conclusions. 

\section{Methods} \label{sec:methods} 

\subsection{General Relativistic Hydrodynamic Simulation} \label{subsec:GRHD}

To model the stellar disruption and the long-term evolution of returning debris around a SMBH, we solve the GRHD equations using the \texttt{HARM3D} code \citep{2009ApJ...692..411N}. The equations solved (expressed in units for which $G = c = 1$) are
\begin{equation} \label{eqn:grmhd}
\begin{aligned}
    \partial_{t}(\sqrt{-g}\rho u^{t}) &= -\partial_{i}(\sqrt{-g}\rho u^{i}), \\
    \partial_{t}(\sqrt{-g}T^{t}_{\;\;\nu}) &= -\partial_{i}(\sqrt{-g} T^{i}_{\;\;\nu}) + \sqrt{-g}T^{\kappa}_{\;\;\lambda}\Gamma^{\lambda}_{\;\;\nu\kappa},
\end{aligned}
\end{equation}
where $g_{\mu\nu}$ denotes the Schwarzschild metric, $\sqrt{-g}$ represents the determinant of the metric, $\rho$ is the total fluid density, $u^{\mu}$ is the four-velocity, and $\Gamma^{\lambda}_{\;\;\nu\kappa}$ denotes the Christoffel symbol, computed using fourth-order finite-differencing on the metric $g_{\mu\nu}$. The stress-energy tensor is given by
\begin{equation}
    T^{\mu\nu} = (\rho + u_{g} + P)u^{\mu}u^{\nu} + Pg^{\mu\nu},
\end{equation}
Here, $u_{g}$ is the total internal energy density and $P$ is the total thermal pressure, both of which account for contributions from both fluid and radiation pressure: $P = \rho k_{B}T/\bar{m} + aT^{4}/3$ and $u_{g} = 3\rho k_{B}T/(2\bar{m})+ aT^{4}$. Here, $k_{B}$ denotes the Boltzmann constant, $\bar{m}=0.62$ is the mean molecular mass in units of the proton mass, and $a$ is the radiation constant. The effective adiabatic index corresponding to this equation of state is \citep{2015ApJ...804...85S}
\begin{equation}
    \Gamma_{\rm eff} = \frac{4 + 5(u_{\rm fluid}/u_{\rm rad})}{3(1 + [u_{\rm fluid}/u_{\rm rad}])}.
\end{equation}
Adiabatic evolution implies no radiation losses. This approximation is justified because the photon diffusion time in our simulation ($\sim 10^{4} - 10^{6}$\,$t_g$, where $t_g = GM_{\rm BH}/c^{3}$ is the gravitational time) is significantly longer than the local dynamical time ($\mathcal{O}(100)\,t_g$) and comparable to the evolution time.  Although radiative cooling is important for accurately estimating observables (as discussed in Section ~\ref{subsec:emobserve}), it is of little importance for overall debris evolution and shock formation. In addition, the equation of state, which accounts for both gas and radiation pressure, assumes local thermal equilibrium. This assumption is also valid because the average thermalization radius ($\simeq 1100 r_g$ at $t=t_0$) is generally well beyond the outermost radius at which shocks take place ($\sim 100 r_g$).

To reconstruct primitive variables at cell boundaries, we employ the piecewise parabolic method \citep{1984JCoPh..54..174C} with a monotonized central-difference slope limiter. The Riemann fluxes are computed using the Lax-Friedrichs solver. Time integration of the GRHD equations is performed using the second-order Strong Stability Preserving Runge-Kutta method \citep{gottlieb2011strong}. We set the Courant-Friedrichs-Lewy number to be $0.7$.

\subsection{Initial Conditions} \label{subsec:initial}

In its initial state, the star to be disrupted is halfway through its main-sequence lifetime, that is, $30\,\%$ of the mass in the star's core is hydrogen. The internal radial profiles of its density and pressure are computed using the stellar evolution code \texttt{MESA} \citep{2011ApJS..192....3P}, assuming solar metallicity. The star has a mass $M_{*} \approx 1$\,$M_{\odot}$, a radius $R_{*} \approx 1\,R_{\odot}$, and a vibration time $\tau_{*} = \left[GM_{*}/(4\pi R_{*}^{3}/3)\right]^{-1/2} \simeq 700$\,$t_g$. The black hole mass $M_{\rm BH} = 10^6\,M_\odot$, making $t_g \simeq 5$\,s. As such, the canonical tidal radius $r_{t} = R_{*}(M_{\rm BH}/M_{*})^{1/3} = 49 $\,$r_g$, although the \textit{physical} tidal radius, the maximum pericenter leading to a full disruption, is smaller by a factor of $\sim 0.5$ \citep{2020ApJ...904...98R} due to stellar structure and relativistic effects.

The star is initially placed on a parabolic orbit around the black hole whose pericenter distance is $r_p = 0.2\,r_t \simeq 9.8\,r_{g}$. For such a deeply penetrating orbit, a complete disruption of the star is expected. The star's initial location is $370$\,$r_g$ from the SMBH, corresponding to $\approx 8$\,$r_{t}$, so that the star reaches the pericenter in a time of approximately $5$\,$\tau_{*}$. We set the origin of the time, $t = 0$, to the moment of the pericenter passage. 

Given the model parameters, the characteristic fallback time of the debris can be estimated as \citep{2020ApJ...904...98R} 
\begin{equation} \label{eqn:tfallback}
\begin{aligned}
    t_{0} = \frac{\pi}{\sqrt{2}}\frac{GM_{\rm BH}}{\Delta \epsilon^{3/2}} &= (7.05 \times 10^5 t_g, 40\,\text{d})~\Xi(M_{*},M_{\rm BH})^{-3/2}\\
    &\times  \left(\frac{M_{\rm BH}}{10^{6}M_{\odot}}\right)^{1/2} \left(\frac{M_{*}}{M_{\odot}}\right)^{-1}\left(\frac{1R_{*}}{1R_{\odot}}\right)^{3/2},
\end{aligned}
\end{equation}
where $\Delta \epsilon$ represents the characteristic spread in the energy distribution
\begin{equation}
    \Delta \epsilon = \frac{GM_{\rm BH}R_{*}}{r_{t}^{2}},
\end{equation}
and the corresponding semi-major axis is $a_0 = GM_{\rm BH}/(2\Delta \epsilon) \simeq 2500$\,$r_g = 3.6\times10^{14}$\,cm. The actual width of the energy distribution is $\Xi \Delta \epsilon$, where $\Xi(M_{*},M_{\rm BH}) \simeq 1.43$ is an order of unity correction factor \citep{2020ApJ...904...98R}. The corresponding timescale $t_0 \simeq 4.1 \times 10^5\,t_g\simeq 23$\,days. We track the evolution up to $1.52$\,$t_{0}$.

The peak mass fallback rate at $t_0$ is
\begin{equation} \label{eqn:mfallback}
\begin{aligned}
    \dot{M}_{0} &= 4 \times 10^{-4} M_{\odot}\textrm{day}^{-1}~\Xi(M_{*},M_{\rm BH})^{-3/2}\\
    &\times  \left(\frac{M_{\rm BH}}{10^{6}M_{\odot}}\right)^{-1/2} \left(\frac{M_{*}}{M_{\odot}}\right)^{2}\left(\frac{R_{*}}{R_{\odot}}\right)^{-3/2}.
\end{aligned}
\end{equation}
%

\subsection{Disruption Phase: Moving Patch Simulation} \label{subsec:single}

The first phase of the calculation is carried out in a moving patch that follows the star's center-of-mass. The term $\Gamma^{\lambda}_{\;\;\nu\kappa}$ contains derivatives of the metric and encapsulates the influence of gravity, incorporating contributions from both the SMBH and the star. Stellar gravity can be treated as a perturbation $h^{\rm sg}_{\mu\nu}$ to the background metric $g^{0}_{\mu\nu}$
\begin{equation}
  g_{\mu\nu} = g^{0}_{\mu\nu} + h^{\rm sg}_{\mu\nu},
\end{equation}
and the post-Newtonian perturbation metric $h^{\rm sg}_{\mu\nu}$ is given by
\begin{equation} 
h^{\rm sg}_{\mu\nu} = \begin{cases}
-2\Phi, & \text{for } \mu = \nu =0 \\
0, & \text{otherwise } 
\end{cases}
\end{equation}
where $\Phi$ is the gravitational potential, obtained by solving the Poisson equation $\nabla^{2} \Phi = 4\pi \rho$. This approach of using solutions to the Poisson equation remains valid only if the background spacetime can be approximated as flat on the stellar scale. However, the background spacetime in the black hole frame is significantly curved. Executing a coordinate transformation to a 3D region traveling with the star's center-of-mass velocity removes much of this curvature, but retains enough that the tidal terms are significant, invalidating a simple implementation of this post-Newtonian scheme. To solve this problem, we compute the star’s self-gravity in a quasi-flat spacetime defined by an orthonormal tetrad formalism \citep{2020ApJ...904...99R}. In this frame, the metric is exactly Minkowski at the origin (the star's center-of-mass) and deviates from the flat spacetime away from the origin due to tidal terms, but they are much weaker. We first transform the metric in the stellar comoving frame with a curved spacetime to the tetrad metric, where the self-gravity component is added, and then transform the metric components back to the stellar comoving frame. 

As the star passes through the pericenter, tidal forces are strong enough to disrupt it, leading to significant elongation. We adaptively expand the simulation domain, enabling us to follow the entire debris evolution using a single domain until its self-gravity becomes weak compared to that of the SMBH\footnote{In \citet{2023ApJ...957...12R}, however, the disruption of the star is followed using \texttt{patchwork} \citep{2024ApJ...974..242A}, which self‑consistently tracks the mass flowing out of the computational domain enclosing the star.}. At the end of the first stage of the calculation, the domain resolution is $(N_{x},N_{y},N_{z}) = (600,600,200)$, with a corresponding cell size of $(dx,dy,dz) = (0.08,0.08,0.04)$, all in units of $r_g$. Outflow boundary conditions are applied across all interfaces. In this first stage, we use a uniform density floor of $6.2\times 10^{-17}$\,g\,cm$^{-3}$ and an internal energy density floor of $5.6\times 10^{-2}$\,erg\,cm$^{-3}$.

We run the first stage of the simulation up to $t = 1400$\,$t_g$. By its end, the center-of-mass position of the debris is approximately $190$\,$r_g$ from the black hole and is moving away from it. The maximum density of the star has decreased by approximately three orders of magnitude compared to its value at the beginning of the simulation. 

\subsection{Fallback Phase: Global Patch Simulation} \label{subsec:two}

Once the stellar self-gravity becomes weak, we interpolate all the fluid variables onto a global Schwarzschild coordinate system, with the SMBH situated at the center of the computational domain. The global coordinates are defined using the so-called `native' coordinate system $(x^{1}, x^{2}, x^{3})$, which can be transformed to standard spherical coordinates via
\begin{equation} \label{eqn:coordinate}
\begin{aligned}
    r &=  \text{exp}(x^{1}), \\
    \theta &= \alpha(\text{tanh}[b(x^{2} - a)] + \tanh[b(x^{2} + a)]) + 0.5\pi, \\
    \phi &= x^{3},
\end{aligned}
\end{equation}
such that
\begin{equation}
\begin{aligned}
    \alpha = -\frac{(0.5\pi - \theta_{0})}{\left[\tanh\left(b(-0.5 - a)\right) + \tanh\left(b(-0.5 + a)\right)\right]},
\end{aligned}
\end{equation}
with $a$ (distinct from the radiation pressure constant) and $b$ being adjustable parameters that control the grid compression toward the midplane. The polar cut-out angle varies between $\theta_{0} = 15 - 30^\circ$ to alleviate timestep constraints near the poles. Even with such a wide polar cut-out, the mass lost through the polar-angle boundaries remains small ($\simeq 2 \%$ of the total stellar mass over the entire simulation). We fix $a = 1.5$ and adaptively tune $b$ between $2.64$ and $6.1$. to ensure that both the horizontal and vertical scale heights are spanned by at least several cells, while maintaining a reasonable timestep. For example, at $t=t_{0}$ the number of cells within one vertical scale height $H$ near the nozzle shock is $\simeq 10$. Here we have defined $H$ through 
\begin{equation} \label{eqn:scale}
\begin{aligned}
  H/r &= \frac{\int\int\rho\sqrt{-g}|\theta - \pi/2|d\theta d\phi}{\int\int\rho\sqrt{-g}d\theta d\phi}.
\end{aligned}
\end{equation}
Within the radial density scale length of the stream $|\rho/(d\rho/dr)|$ at the same location, there are several tens of cells.

We adjust the $r$ and $\phi$ extents to fully enclose the stellar debris. The inner radial boundary $r_{\rm in}$ starts out at $5$\,$r_g$, and is moved to $7$\,$r_g$ at $0.3$\,$t_0$ to alleviate the timestep constraint near the black hole, which arises from hot fluid with large sound speed settling into high angular velocity orbits. The outer radial boundary starts out at $5000$\,$r_g$, and is gradually moved to $14000$\,$r_g$ in order to ensure that the bound mass returning within $5$\,$t_{0}$ remains entirely inside the domain. The computational box comprises $800 \times 128 \times 800$ cells and is fixed throughout the simulation.

Prior to the first pericenter passage of the debris, when the initial self-intersection shock occurs due to relativistic precession, we enforce exact entropy conservation during the inversion from the conservative to primitive variables. We then switch to the energy-conserving scheme right before the first self-intersection shock appears in order to capture the irreversible heating produced by shock interactions. Outflow boundary conditions are applied at the $r$ and $\theta$ interfaces, while periodic boundary conditions are imposed along the $\phi$ direction. 

We employ a radially dependent floor for the density ($\rho_{\rm fl}$) and internal energy density ($u_{\rm fl}$):
\begin{equation}
\begin{aligned}
    \rho_{\rm fl} &= 6.2\times10^{-20}(\frac{r}{50\,r_g})^{-3}\,\textrm{g\,cm$^{-3}$}, \\
    u_{\rm fl} &= 5.6\times10^{-5}\text{MAX}(\frac{r}{50\,r_g}, 2.5)^{-4}\,\textrm{erg\,cm$^{-3}$}.
\end{aligned}
\end{equation}
%

\section{Results} \label{sec:results} 

\subsection{Overview} \label{subsec:overview}
\begin{figure*}[htb!]
    \centering
    \includegraphics[width=1.0\linewidth]{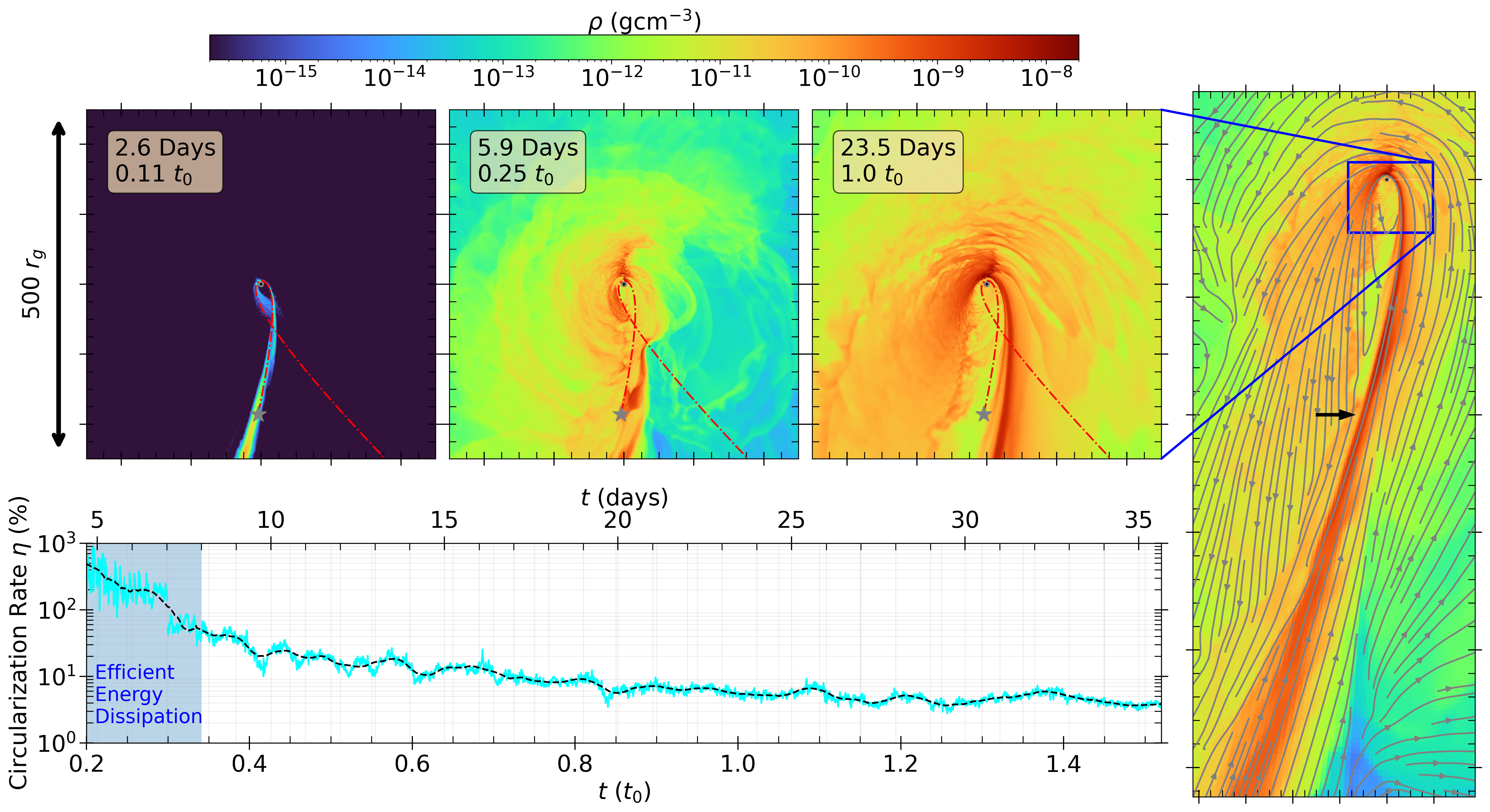} 
    \caption{The overall debris evolution (top) in our simulation, along with the circularization rate (bottom). The top row shows the density contours in the $x-y$ plane, spanning a distance of $500$\,$r_g$, at three different times, representing the early phase (left), its transition (middle) to the late phase (right). The red dash-dotted line traces the expected ballistic geodesic of the incoming debris, and the grey star-shaped scatter marks the initial position of the star. The vertical long panel on the right shows a zoomed-out view of the density at $23$\,days, with velocity streamlines (grey) appended. We include a dark arrow marking a representative location of self-intersection shocks (see \S~\ref{subsec:shock}). The velocity fields are the three-velocity of the fluid, defined as $V^{i} = \sqrt{g_{ii}}u^{i}/u^{t}$, where $g_{ii}$ is the $(i,i)$ metric component and $u^{\mu}$ is the four-velocity. The circularization rate, defined in Equation~\ref{eqn:eff}, measures the pace of circularization: $100$\,\% for complete circularization within $1$\,$t_{0}\simeq 23$\,days (see \S~\ref{subsec:circular}). The dark dashed line represents the moving average of the time series. In the bottom row, the blue shaded region marks the regime where relativistic effects are strong and promote energy dissipation via shocks. \label{fig:one_figure_summary}}
\end{figure*}

The key finding of this work is that \textit{strong relativistic effects do not significantly facilitate swift formation of an accretion disk, making the overall debris evolution qualitatively similar to that of weakly relativistic TDEs}. This finding contrasts with the conventional expectation for strongly relativistic TDEs, in which rapid energy dissipation via strong shocks occurring at $r \sim r_p$, driven by general relativistic effects, is thought to produce an accretion disk on a timescale comparable to the peak mass-return time ($t_0\simeq 23$\,days in our simulation).

At the onset of the debris fallback, shocks, driven by compression at the pericenter and by stream-stream intersections at larger distances, are moderately strong, dissipating orbital energy at a rate $\sim (3-5) \times 10^{-3}\dot{M}c^2$, where $\dot{M}$ is the mass fallback rate (see top-left and top-middle panels of Figure~\ref{fig:one_figure_summary}), although this dissipation rate per unit mass is at least an order of magnitude smaller than what is generally expected. 

However, self-intersection shocks, in which the angle between the two streams is large, push the incoming debris sideways, giving the incoming debris greater angular momentum and therefore a larger pericenter distance. Consequently, when the incoming stream passes through the pericenter, its apsidal precession is weaker (see the top-middle and top-right panels of Figure~\ref{fig:one_figure_summary}).

As a result, the shocks suffered by the incoming debris steadily weaken until, by $t \simeq 0.3\,t_0$, their dissipation rate levels out at a value comparable to that of weakly relativistic TDEs. Because circularization depends directly on dissipation, its progress also slows down. If one defines the `circularization efficiency' as the heating rate per $t_0$, normalized by the binding energy of a circular orbit of radius $2\,r_p$, it drops from over $100$\,\% to only a few percent (see the bottom panel of Figure~\ref{fig:one_figure_summary} and further discussion in \S~\ref{subsec:circularization}). From $t \simeq 0.3\,t_0$ until our simulation ended at $t \simeq 1.52\,t_0$, the event evolves very much like those previously studied, i.e., weakly relativistic TDEs, with larger pericenter distances.

Consequently, even at $\simeq 1.52$\,$t_0\simeq 35$\,days, the debris remains eccentric (eccentricity $e \gtrsim 0.4 - 0.8$, see also the zoomed-out panel of Figure~\ref{fig:one_figure_summary}), and only $\mathcal{O}(10^{-3})\,M_\odot$ has been accreted onto the black hole. Most of the returned mass stays around the apocenter.

There is, however, one way in which these relativistic TDEs differ from the ordinary case. During the disruption of the star, the debris' energy spreads substantially wider than predicted by the conventional impulse approximation; in fact, between $-2E/\Delta\varepsilon$ and $-4E/\Delta\varepsilon$, the energy distribution's $e$-folding length is $\simeq 1.5\times$ greater than in the energy distribution found for $M_* = 1\,M_\odot$ in \citet{2020ApJ...904..101R}. As a result, the fallback rate rises rapidly, achieving a substantial rate by $0.25$\,$t_0$ and then forming a plateau that lasts for several $t_0$ (a few months for our simulation parameters). This behavior contrasts with the usual sharp peak at $t \simeq t_0$, followed by a $t^{-5/3}$ decay (see the left panel of Figure~\ref{fig:combined_distribution} for reference).

\subsection{Stellar Disruption Phase} \label{subsec:disruption}
\begin{figure*}[htb!]
    \centering
    \includegraphics[width=1.0\linewidth]{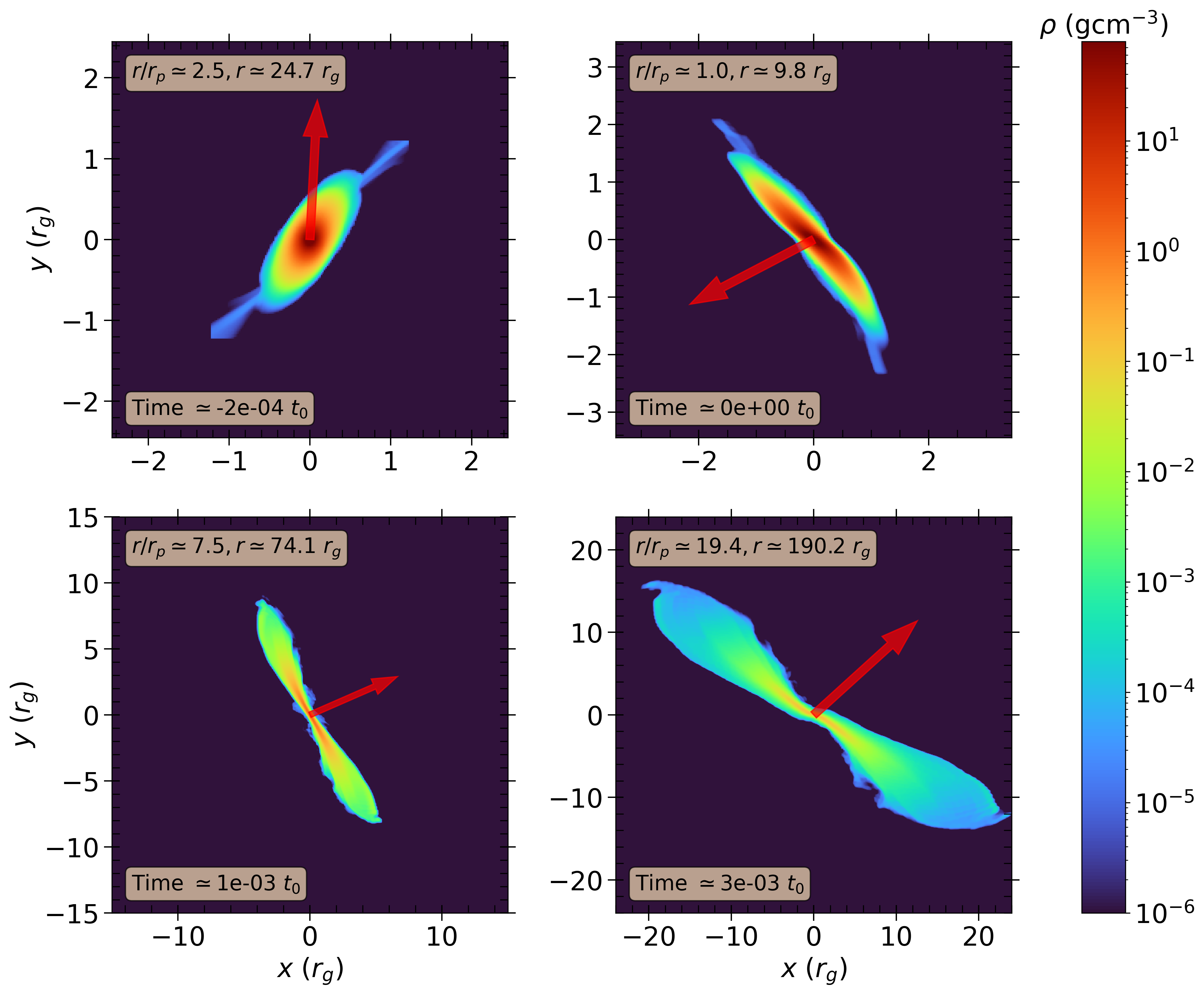} 
    \caption{Temporal evolution of the density contour (in the log$_{10}$ scale) along the equatorial plane in the domain comoving with the center-of-mass of the star. Note that the plotting extent increases with time. In each panel, the upper-left text box lists the position of the center-of-mass of the fluid in units of $r_g$ and $r_p$; the lower-left text box indicates the time since the pericenter passage in $t_0$. In each panel, the red arrow indicates the direction toward the SMBH. \label{fig:rho_stage1_xy}}
\end{figure*}
\begin{figure*}[htb!]
    \centering
    \includegraphics[width=1.0\linewidth]{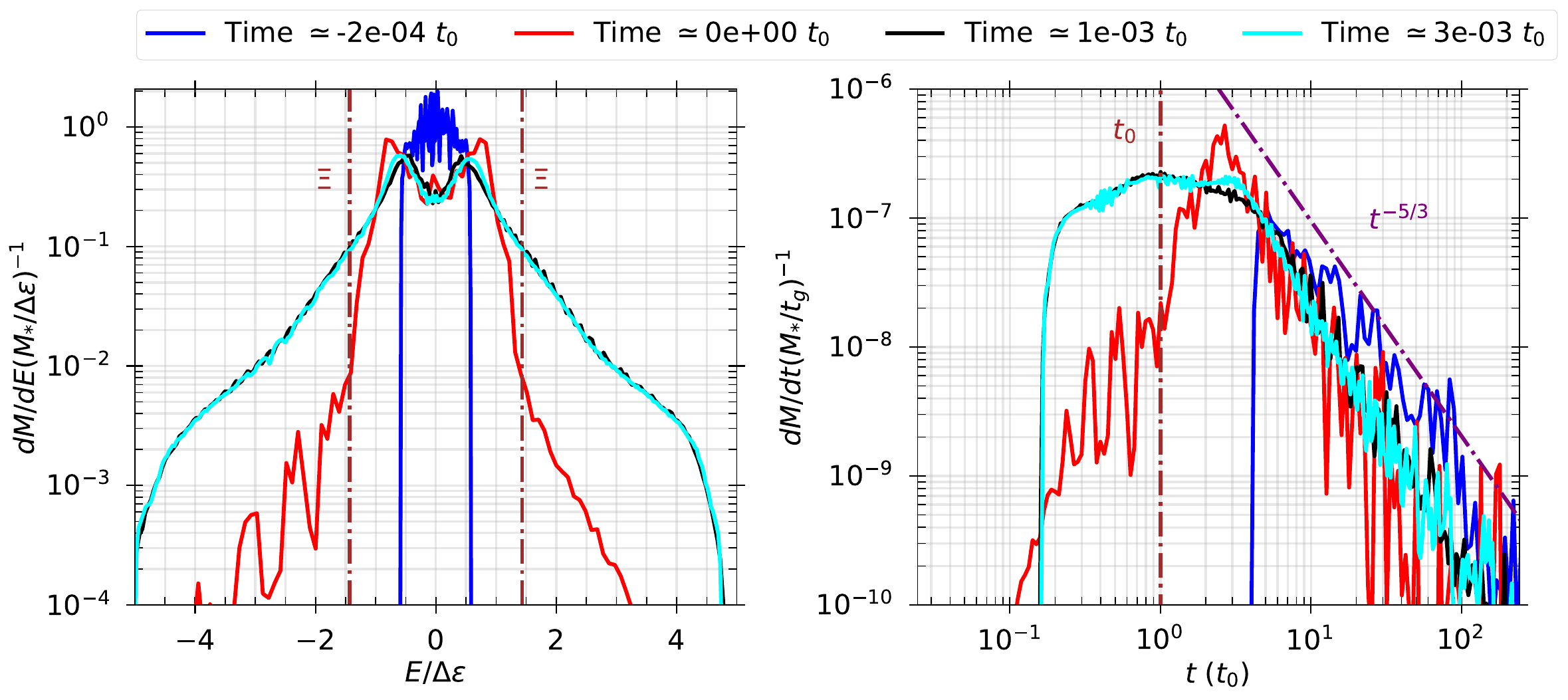} 
    \caption{Orbital energy (left) and fallback rate (right) distributions of the stellar debris at a few different times. In the left-hand panel, we include vertical brown lines that indicate the $\Xi$ factor from Equations~\ref{eqn:tfallback} and~\ref{eqn:mfallback}. In the right-hand panel, we indicate the peak fallback time $t_0$ (brown dashed-dotted line) and the scaling relation $t^{-5/3}$ (purple dashed-dotted line) for reference. \label{fig:combined_distribution}}
\end{figure*}
\begin{figure*}[htb!]
    \centering
    \includegraphics[width=0.48\linewidth]{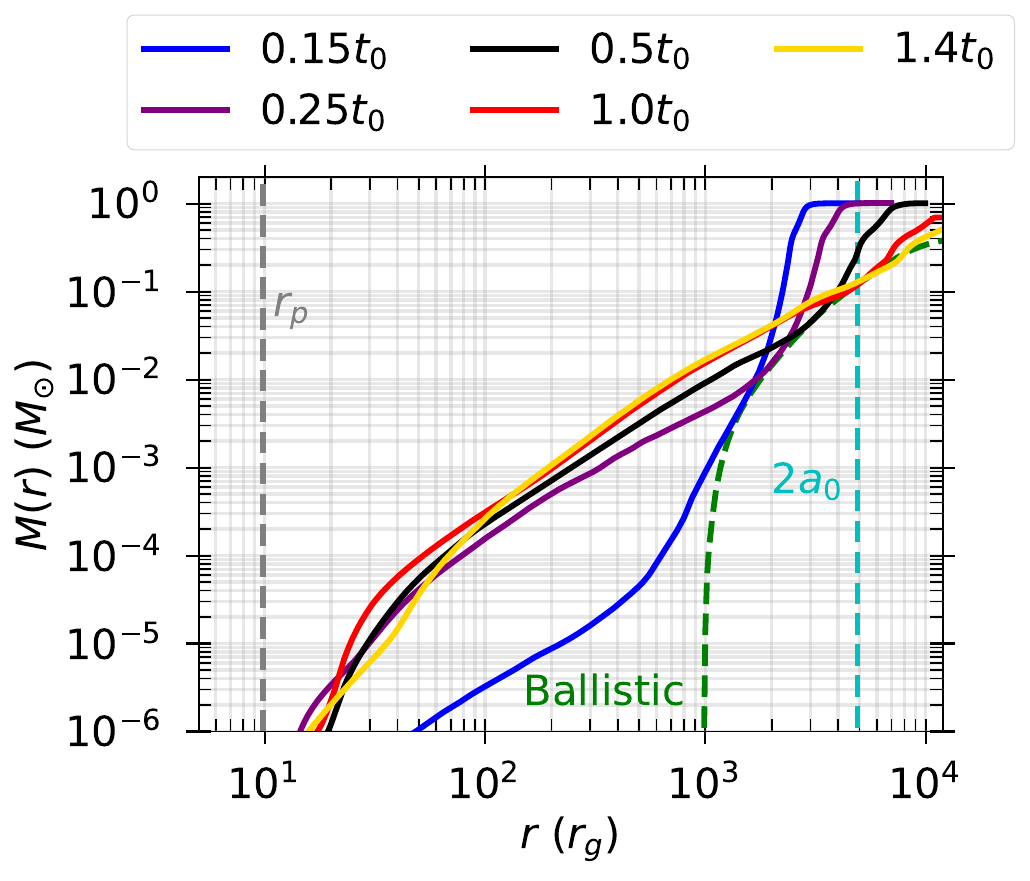} 
    \includegraphics[width=0.48\linewidth]{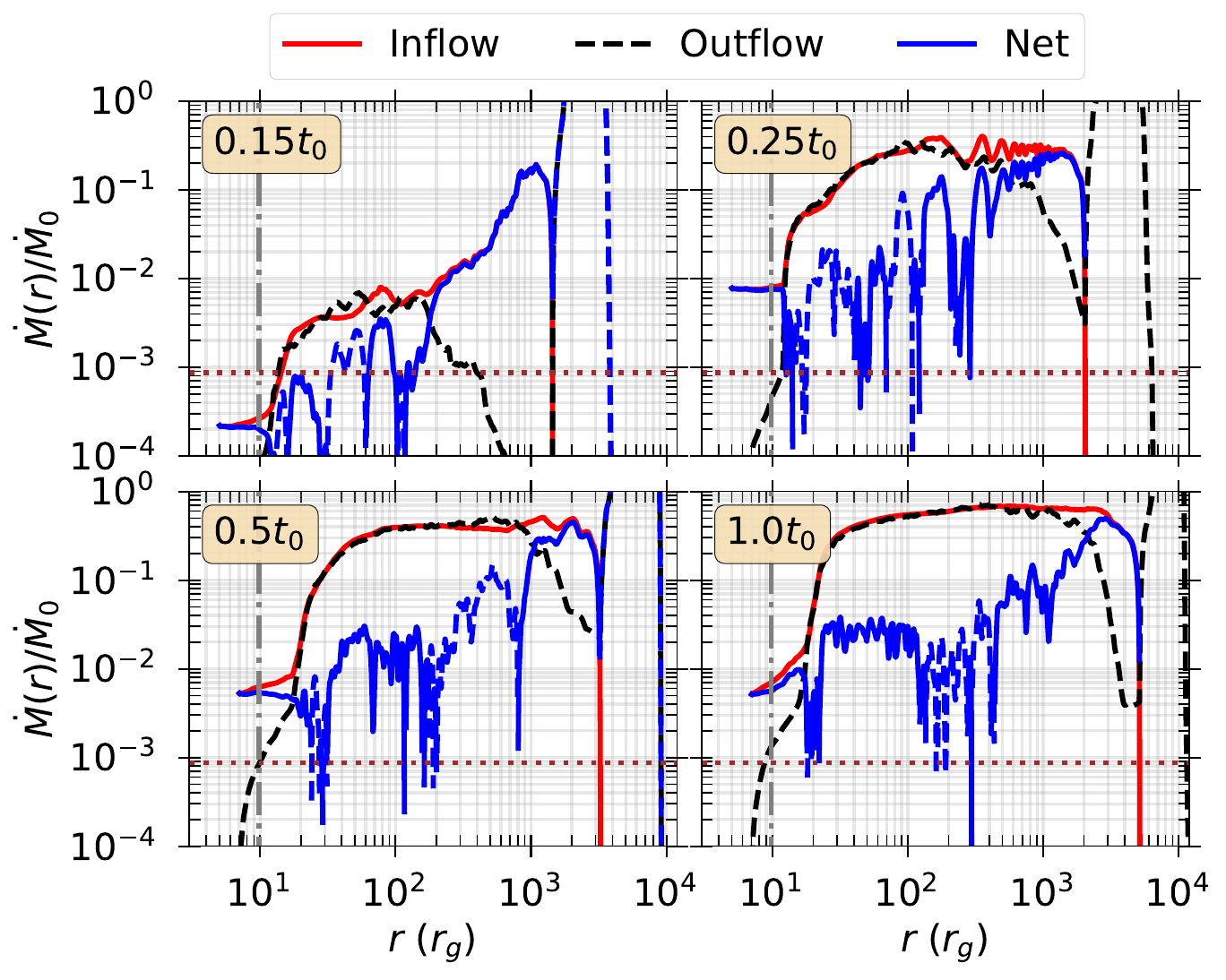} 
    \caption{(Left) Time evolution of the enclosed mass profile at five epochs. The green dashed curve shows the mass whose apocenter $\approx 2a(E) < r$ for the $dM/dE$ shown in the left panel of Figure~\ref{fig:combined_distribution}. Note that it is hidden under the other curves for $r \gtrsim 1100\,r_g$. (Right) Profiles of the mass inflow/outflow rate, normalized to the canonical peak mass-return time $\dot{M}_0$, at the first four times shown in the left panel. Contributions from outgoing (dashed black) and incoming (solid red) fluid, and their net (blue) are shown separately. For the net rate, we use different line styles to distinguish net inflow (solid) and outflow (dashed) rates. In both panels, the vertical grey lines indicate the initial stellar pericenter distance, while for the right panel only, the horizontal brown line indicates the Eddington mass accretion rate, assuming a radiative efficiency of unity. \label{fig:mrmdot}}
\end{figure*}

Figure~\ref{fig:rho_stage1_xy} shows the density contours at several epochs during the disruption phase. At $t \approx -100$\,$t_g$ (top-left), the star, located at $\simeq 2.5$\,$r_p \simeq 0.54$\,$r_{t} \simeq 25\,r_g$ from the SMBH before its pericenter passage, has already been deformed by strong tidal forces. These forces also exert a torque on the deformed star, causing it to rotate with respect to its center-of-mass, but the dense stellar core (deep red) remains roughly spherical. The star is stretched further as it approaches the pericenter at $t \approx 0.0$\,$t_g$ (top-right). At this point, the densest material of the star begins to be elongated. At $t \approx 400$\,$t_g$ (bottom-left), the star is highly elongated and increasingly aligned toward the SMBH due to ongoing tidal torques. At $t \approx 1400$\,$t_g$ (bottom-right), when the star's center-of-mass has reached $20$\,$r_p \simeq 4$\,$r_{t}$, the debris continues to expand, but at a slower rate. The maximum density at $t = 1400$\,$t_g$ is three orders of magnitude lower than that of the undisturbed star ($\sim 150$\,g\,cm$^{-3}$). At this point, the tidal gravity exceeds the stellar self-gravity across the domain by at least $\mathcal{O}(10^{2})$.

We show in the left panel of Figure~\ref{fig:combined_distribution} the orbital energy distribution $(E = -u_t - 1)$ of the debris at the same epochs as in Figure~\ref{fig:rho_stage1_xy}. At $t \approx -100$\,$t_g$, the distribution peaks near $E = 0$. Between $t \approx -100$ and $0$\,$t_g$, the energy distribution is broadened, but not significantly. After the pericenter passage, it broadens substantially until $t \approx 400$\,$t_g$, when the star's center-of-mass is at $r\simeq 74$\,$r_g$. By this time the distribution has developed pronounced shoulders and extended wings that evolve very little thereafter. This result is qualitatively consistent with previous TDE simulations accounting for realistic stellar structure \citep[e.g.,][]{2019MNRAS.487..981G,2019ApJ...882L..25L,2020ApJ...904...98R}.

The corresponding fallback curves for the bound debris at the same epochs are illustrated in the right panel of Figure~\ref{fig:combined_distribution}. We compute the fallback rate as $dM/dt = (dM/dE)|dE/dt|$, where $dM/dE$ is the energy distribution shown in Figure~\ref{fig:combined_distribution} and $|dE/dt|$ is calculated assuming Keplerian orbits. As expected, the most significant changes occur during the pericenter passage, between $-100$\,$t_g$ and $400$\,$t_g$. Afterward, the mass-return rate curve freezes, showing a sharp rise at $t \simeq 7 \times 10^{4}$\,$t_g$, a plateau that extends to $10^{6}$\,$t_g$, and a subsequent $t^{-5/3}$ decline. From this curve, we identify the peak fallback time as $\simeq 4.11\times10^{5}$\,$t_g$, almost exactly at $t_0$. It is marked by the vertical dot-dashed brown line in the right panel. The corresponding peak rate of $\simeq 600\,L_{\rm Edd}/c^{2}=600\,\dot{M}_{\rm Edd}$ (assuming the radiative efficiency to be $1$) agrees with the nominal estimate $M_{*}/[3t_{0}]$. The sharp rise and plateau phase reflect the broad wings of the energy distribution.

From now on, we focus on the bound debris that returns to the SMBH through the end of the simulation ($1.52\,t_{0}\simeq 35$\,days).

\subsection{Mass and Energy Evolution of the Bound Debris} \label{subsec:fallback}
\begin{figure}[htb!]
    \centering
    \includegraphics[width=1.0\linewidth]{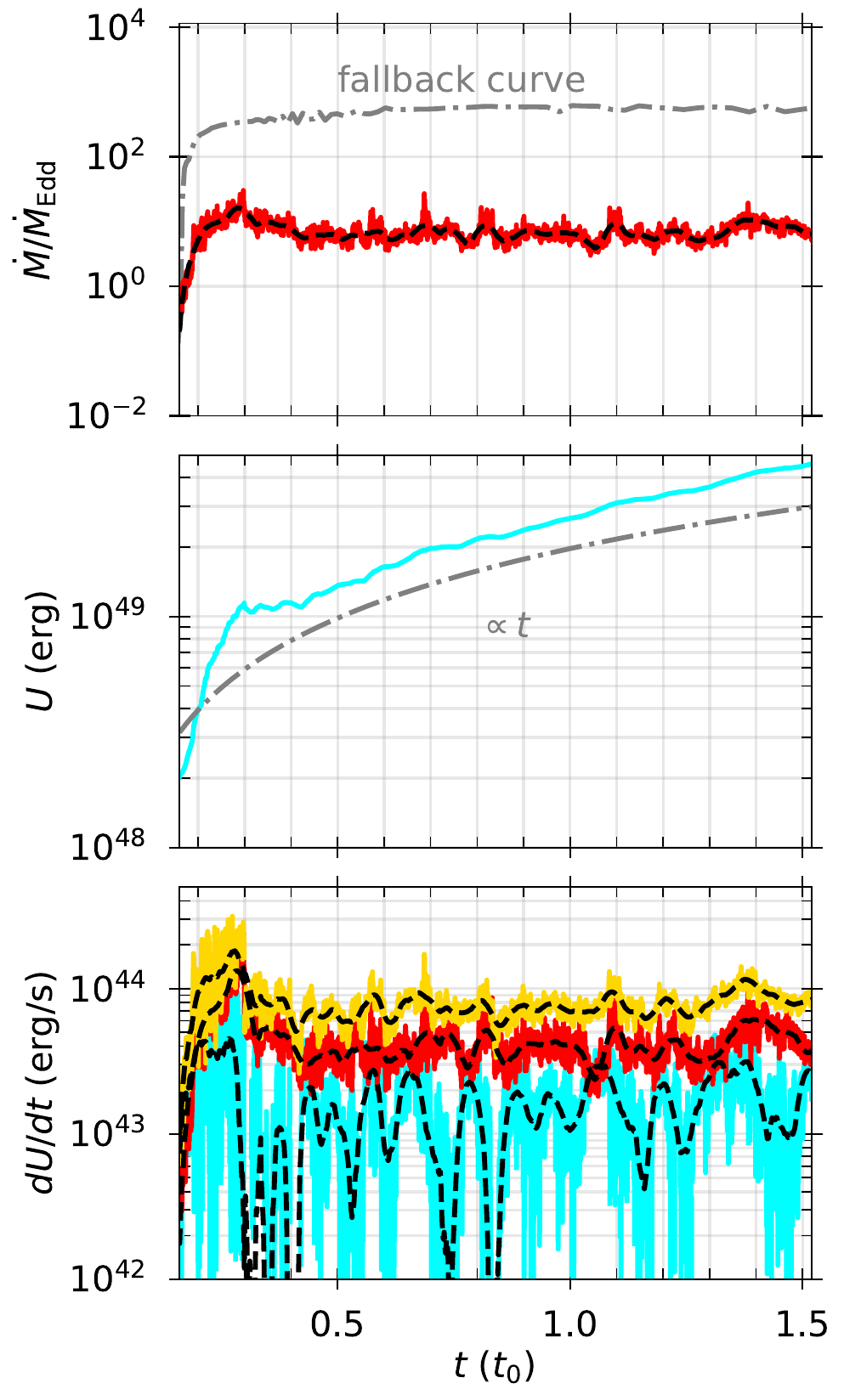} 
    \caption{Time evolution of the mass accretion rate (top), the total internal energy within the simulation domain (middle), and the rates of internal energy change (bottom). In the top panel, the mass accretion rate is normalized by the Eddington value (assuming a radiative efficiency of unity) and the gray dot-dashed line shows the (similarly normalized) mass fallback rate, when it reaches a steady state, which is also shown in the right panel of Figure~\ref{fig:combined_distribution}. In the bottom panel, we compare: the heating rate, i.e., internal energy changes by means of shocks or adiabatic compression/expansion (yellow); the advected internal energy loss rate, i.e., losses due to mass flows across the inner boundary (red); and the net rate of internal energy change within the domain (cyan), i.e., the difference between the first two minus the energy carried out of the domain when gas leaves through the $\theta$ boundary. In the top and bottom panels, a black dashed line represents the time-averaged value of the corresponding time series. In the middle panel, we include a $\propto t$ scaling relation. \label{fig:time_series_summary}}
\end{figure}

\subsubsection{Mass and energy of the debris within the domain}
We present in the left panel of Figure~\ref{fig:mrmdot} the time evolution of the enclosed mass distribution $M(<r)$, defined as
\begin{equation}
    M(r) = \int_{r_{\rm in}}^{r}\int\int\sqrt{-g}\rho u^{t}dx^1dx^2dx^3.
\end{equation}
Also shown in the same figure (green dashed curve) is the same quantity if all fluid were located at $2a(E)$, where $a(E)$ is the orbital semi-major axis and the fluid followed the energy distribution $dM/dE$ shown in Figure~\ref{fig:combined_distribution}. We also include a vertical cyan line marking the characteristic apocenter distance of $2\,a_0$ (see \S~\ref{subsec:initial}), which corresponds to the apocenter of highly eccentric debris with an orbital period of $t_0$. At early times, more mass is found at $r \gtrsim 2000\,r_g$ than would be predicted by placing it all at its apocenter; however, when $t \gtrsim 0.5 - 1.0\,t_0$, the latter approximation is very good at these large radii. There is always more mass at $r \lesssim 2000\,r_g$ than the `apocenter-only' approximation would give, but in absolute terms this mass is a very small fraction of $M_*$. The large-radius excess at early times is due to the presence of low-energy unbound matter, which leaves the domain by $\sim 0.5 - 1.0\,t_0$. The small amount of mass closer to the black hole is mostly determined by the fraction of an orbital time the fluid spends at $r < 2a(E)$. Thus, we see that for the first $1.52\,t_0$, the orbital energy of the overwhelming majority of the mass changes very little; that is to say, hardly any of it has `circularized' to form a disk. If the returned mass largely follows its ballistic motion, it implies that the fluid is mechanically rather than thermally supported.

To display the overall flow of the debris, we show in the right panel of Figure~\ref{fig:mrmdot} the mass-flow rates within the domain, computed as
\begin{equation} \label{eqn:mdotr}
    \dot{M}_r(r) = -\int\int\rho u^{r}\sqrt{-g}dx^{2}dx^{3}.
\end{equation}
At any given time, the radial profile of the mass flux can be divided into four regimes: \textbf{1)} a radially constant inflow, \textbf{2)} nearly comparable inflow and outflow, \textbf{3)} dominantly inflow, and \textbf{4)} dominantly outflow. The radial extent of each regime varies over time. We focus on an example snapshot at $t \simeq 1\,t_{0}$: the net mass flow within $r\simeq 20\,r_{g}$ is inward, and it drops by only a factor of a few all the way to the inner radial boundary ($r = 7\,r_{g})$. At $r \simeq 20 - 3\times10^{3}\,r_{g}$, the inflow and outflow rates are comparable, giving the magnitude of the net flow a value much smaller than that of either component individually, and the sign of the net flow fluctuates. At $r\simeq 3\times10^{3} - 5\times 10^{3}\,r_{g}$, the mass flux is dominantly inward, driven by debris returning from radii beyond the maximum apocenter distance of the material that has already returned. The dominantly outward flux at $r\gtrsim 5\times 10^{3}\,r_{g}$ corresponds to both the loosely bound debris still moving outward and unbound debris. The inward flux within $r\simeq 10 - 20\,r_{g}$ diminishes the mass in that region, as seen at $t\simeq 0.25 - 1.4\,t_{0}$ in the left panel of Figure~\ref{fig:mrmdot}. Meanwhile, the nearly constant enclosed mass over time between $20\,r_g$ and $3\times10^{3}\,r_{g}$ can be explained by the very small net mass flux through that region. 

In the middle panel of Figure~\ref{fig:time_series_summary}, we show the time evolution of $U$, the debris' total internal energy within the computational domain. It is computed as 
\begin{equation} 
    U = -\int [(u_g + P)u^tu_t + P] \sqrt{-g}dx^{1} dx^{2} dx^{3}.
\end{equation}
We find that $U$ increases rapidly from $2\times 10^{48}$\,erg at $0.15\,t_0$ to $10^{49}$\,erg at $0.3\,t_{0}$, followed by a relatively slower rise to $4\times 10^{49}$\,erg at $t\simeq 1.5\,t_{0}$.

\subsubsection{Accreted mass and energy}

The mass accretion rate onto the SMBH can be estimated by the mass flux through the inner radial boundary ($5$\,$r_{g}$ before $0.3$\,$t_{0}$ and $7$\,$r_{g}$ after $0.3$\,$t_{0}$). As shown in the upper panel of Figure~\ref{fig:time_series_summary}, the accretion rate rises sharply during the first $0.3$\,$t_{0}$ to $\sim 10\,\dot{M}_{\rm Edd}$, and then plateaus at that level until the end of the simulation. The total accreted mass up to $1.5\,t_{0}$ is $\simeq10^{-3}$\,$M_{\odot}$, amounting to only about $0.2$\,\% of the total bound mass. Although its time-dependence resembles the mass fallback rate (gray), the accretion rate is approximately two orders of magnitude lower. This contrast implies that all but a fraction $\lesssim 10^{-2}$ of the returning mass passes through the pericenter region without being accreted and then heads outward again. 

The accreted mass carries thermal energy which would otherwise be radiated. The internal energy accretion rate is evaluated at $r = r_{\rm in}$ as
\begin{equation}
\begin{aligned}
    \dot{U}_{\rm in} & = - \int\int[(u_g + P)u^ru_t]\sqrt{-g}dx^{2}dx^{3}|_{r=r_{\rm in}},
\end{aligned}
\end{equation}
and is depicted in the bottom panel of Figure~\ref{fig:time_series_summary} as the red line. The overall shape of $\dot{U}_{\rm in}$ closely tracks that of $\dot{M}_r$, showing an initial outburst followed by a plateau at a somewhat lower level. Interestingly, the initial peak is near-Eddington ($\simeq 10^{44}$\,$\mathrm{erg\,s^{-1}}$), while the plateau is a factor of several lower $(3-4\times 10^{43}\,\mathrm{erg\,s^{-1}}$). The gross thermal energy generation rate (yellow curve in the same panel) is obtained by adding the energy escape rate through all boundaries to the partial time derivative of $U$ (cyan curve). The sub-Eddington-level energy accretion rate---roughly $10$\,\% of the gross thermal energy generation rate---indicates that a considerable amount of thermal energy is accreted ($\simeq 10^{49} - 10^{50}$\,erg), rather than radiated. 

As illustrated in the bottom panel of Figure~\ref{fig:time_series_summary}, the rate at which $U$ increases due to dissipative heating and the net of adiabatic compression and expansion is always greater than the rate at which $U$ decreases by advection out of the domain. However, the contrast is small, generally only a few tens of percent. This coincidence means that the net rate of increase for $U$ (as shown by the cyan curve in Figure~\ref{fig:time_series_summary}) is slower by about an order of magnitude than the total heating rate. 

From $t \simeq 0.3\,t_0$ onward, there is essentially no long-term trend in the total heating rate, i.e., it plateaus and fluctuates around $\sim \mathcal{O}(10^{44})$\,erg\,s$^{-1}$. This behavior is consistent with the hydrodynamics we have described: initially strong shocks that gradually weaken and eventually level out in strength.

\subsection{Shocks} \label{subsec:shock}
\begin{figure*}[htb!]
    \centering
    \includegraphics[width=1.0\linewidth]{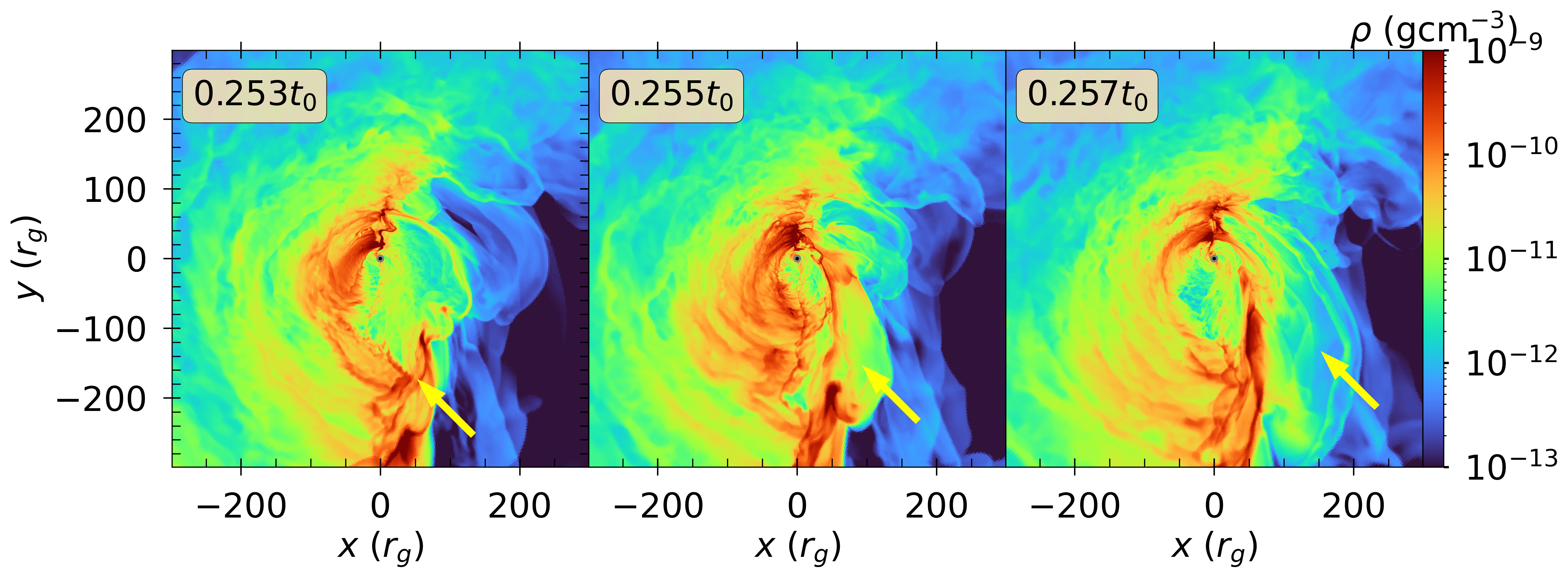} 
    \caption{Simulation snapshots of the fluid density along the $x-y$ plane at selected epochs (indicated in the upper-left text box), illustrating early-time violent self-intersection shocks that are strong enough to significantly disrupt the incoming streams (see also \citet{2022MNRAS.510.1627A}). In the left panel, the yellow arrow marks the rough location of the self-intersections, while in the middle and right panels, the arrow indicates the perturbed debris due to the stream collisions. \label{fig:violent_streams_temp}}
\end{figure*}
\begin{figure}[htb!]
    \centering
    \includegraphics[width=1.0\linewidth]{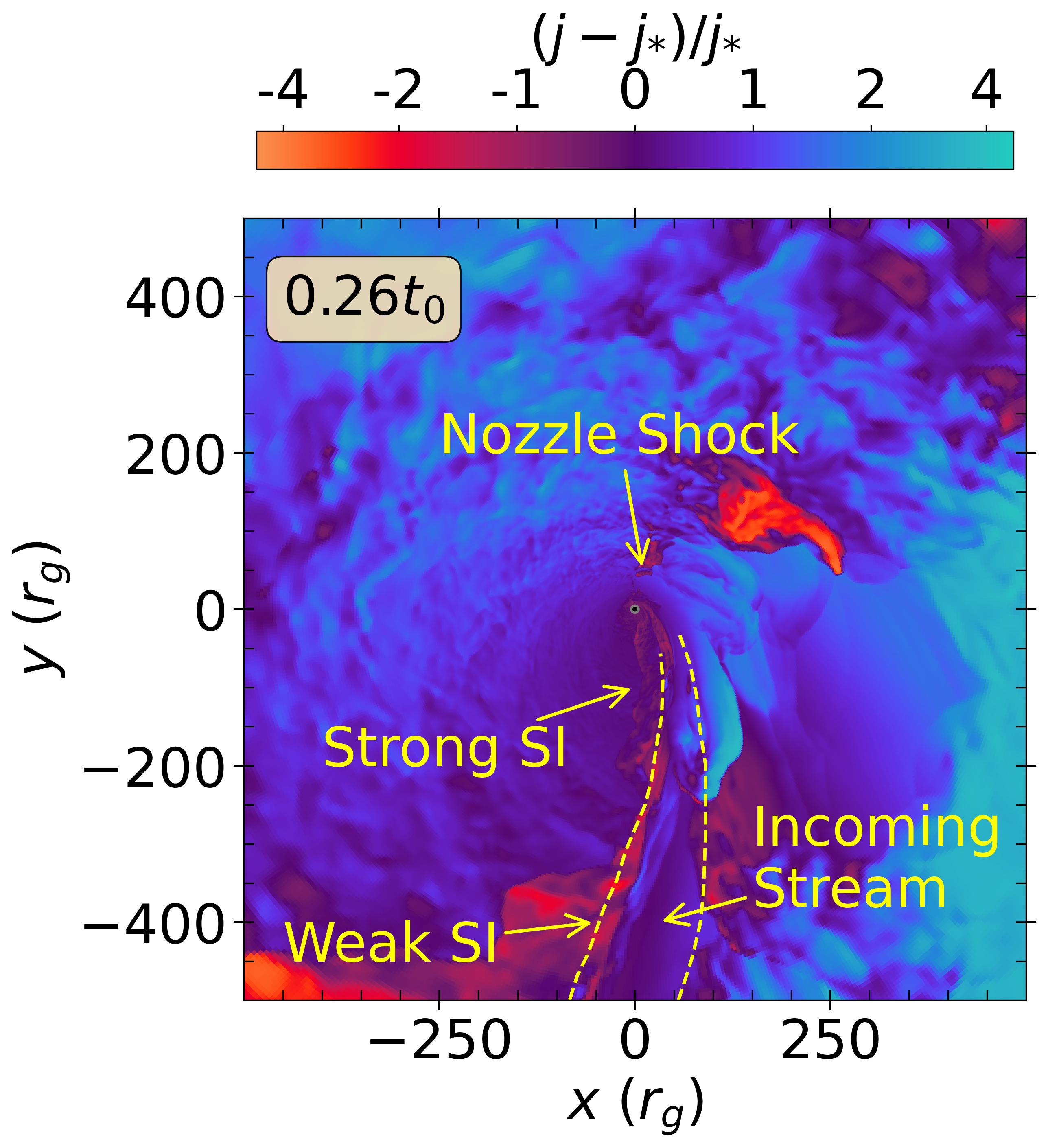} 
    \caption{Fractional specific angular momentum relative to the star’s initial value, shown along the $x-y$ plane at $t\simeq 0.26$\,$t_0$ (which corresponds to the middle panel of Figure~\ref{fig:violent_streams_temp}). We approximately mark the boundaries of the incoming stream and indicate the positions of the nozzle shock, as well as strong and weak self-intersection (SI) shocks. \label{fig:eorb_nozzle_0.15t0}}
\end{figure}
\begin{figure*}[htb!]
    \centering
    \includegraphics[width=1.0\linewidth]{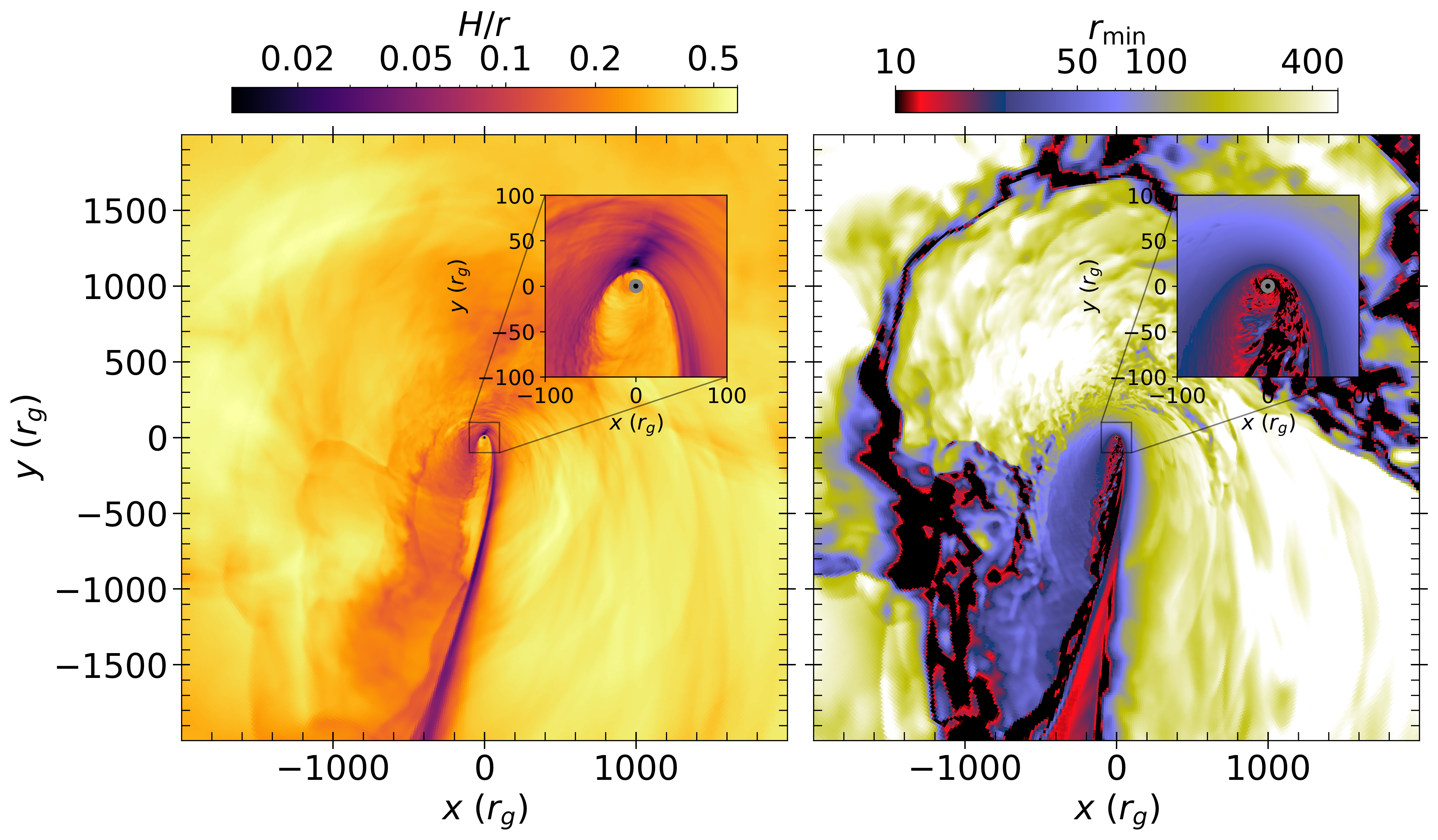} 
    \caption{Contours of the local vertical scale height (see Equaton~\ref{eqn:scale}) and the Newtonian orbital pericenter (right) in the $x - y$ plane at $1$\,$t_0$. Each panel includes a zoomed-in view of the region near the SMBH. As discussed in the main text, the pericenter distance of the incoming debris increases due to angular-momentum transfer from the outgoing debris. This feature persists even in the late phase. In the Newtonian pericenter ($r_{\mathrm{min}}$) map, we illustrate this clearly by showing a transition zone from small $r_{\mathrm{min}}$ (red) to large $r_{\mathrm{min}}$ (purple) after the incoming gas passes through the self-intersection shocks. \label{fig:1t0_shock}}
\end{figure*}
\begin{figure}[htb!]
    \centering
    \includegraphics[width=1.0\linewidth]{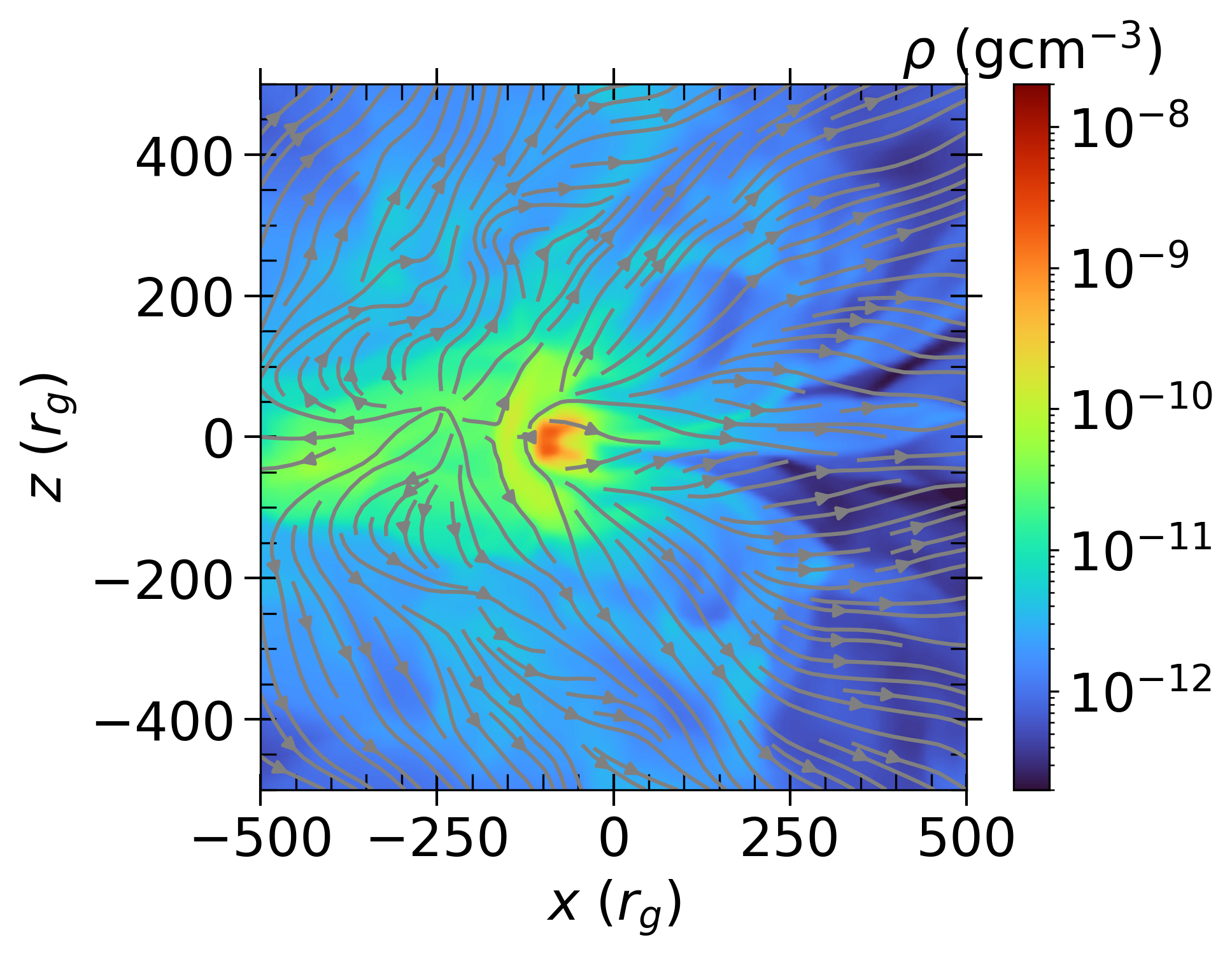} 
    \caption{Snapshots of the fluid density in the $x-z$ plane at $1$\,$t_0$, overlaid with the velocity streamlines (grey). The snapshot is taken at $y = -1000$\,$r_g$, corresponding to the location indicated by the dark arrow of the zoomed-out view in Figure~\ref{fig:one_figure_summary}. \label{fig:weak_stream_1t0}}
\end{figure}

The returning stellar debris undergoes two types of shocks: a `nozzle' shock, which arises from the vertical convergence of the fluid toward the mid-plane near the orbital pericenter; and self-intersection shocks, which result from the crossings between the incoming and outgoing streams. In the early phase of the simulation, these shocks, driven primarily by a small pericenter distance and strong relativistic apsidal precession\footnote{For the self-intersection shocks, there are also contributions from the apsidal rotation due to the finite duration of the disruption, but they become more important than relativistic precession only for $r_p \gtrsim 30\,r_g$; see \citet{2015ApJ...804...85S}}, dissipate the orbital energy of the debris and efficiently circularize the gas's orbits. As the simulation progresses, however, the energy dissipation rate diminishes, making the circularization of the debris less effective. We will describe this transition in more detail below.

\subsubsection{Strong shocks at early times ($t\lesssim 0.3\,t_{0}$)}
For a short time at the beginning of the event, the returning streams undergo a nozzle shock near the pericenter at approximately the initial $r_{p}$ (the top-left panel of Figure~\ref{fig:one_figure_summary})\footnote{The shock location is shifted by $120^{\circ}$ from the initial pericenter along the $x-$axis, due to two $60^{\circ}$ apsidal precessions during the first stellar passage and the subsequent debris passage.}. Adiabatic heating due to vertical compression and dissipation of orbital energy into heat at the nozzle shock combine to raise the debris temperature from below $10^4$\,$\mathrm{K}$ to $\simeq 10^6$\,$\mathrm{K}$.

Subsequently, the outward-moving shocked stream collides with newly incoming debris, producing self-intersection shocks. Strong apsidal precession causes these encounters to occur close to the initial stellar pericenter, where streams with high velocities collide and generate violent shocks that significantly perturb the incoming stream (Figure~\ref{fig:violent_streams_temp}, see also \citealt{2022MNRAS.510.1627A}). These interactions produce intense self-intersection shocks (arrows marking the `Strong SI' region in Figure~\ref{fig:eorb_nozzle_0.15t0}).

During these encounters, the incoming debris is displaced from its expected orbit and gains angular momentum, often enough to raise $j$ by a factor $\sim O(1)$. The blue regions in the specific angular momentum map of Figure~\ref{fig:eorb_nozzle_0.15t0} show gas that has gained angular momentum by this means. Correspondingly, some of the outgoing debris loses angular momentum (colored red in Figure~\ref{fig:eorb_nozzle_0.15t0}), and can lose enough to be deflected onto retrograde orbits. 

The incoming stream, having gained angular momentum, now passes through a pericenter with a distance larger than the star's. As a result, it experiences a weaker nozzle shock, and, when it turns outbound, weaker self-intersection shocks as well. These weaker shocks only mildly perturb subsequent debris (arrows indicating `Weak SI' in Figure~\ref{fig:eorb_nozzle_0.15t0}), thereby allowing the later-arriving debris to undergo violent self-intersections. This strong-weak shock cycle continues until $t\simeq 0.3$\,$t_0$, driving variability in the volumetric energy dissipation rate (yellow line in the bottom panel of Figure~\ref{fig:time_series_summary}), which rises to $10^{44}$\, $\mathrm{erg\,s^{-1}}$. Despite this modulation, the mean dissipation rate is high enough to give the gas returning this early a binding energy corresponding to an orbit with semi-major axis $\simeq 60\,r_g$. 

\subsubsection{Transition to weaker shocks ($t\gtrsim 0.3\,t_{0}$)}
At later times, the dissipation rate plateaus at $\lesssim 10^{44}$\,$\mathrm{erg\,s^{\mathnormal{-1}}}$ (middle panel of Figure~\ref{fig:time_series_summary}), indicating that the shocks have weakened. The shocks weaken via two mechanisms: continuing angular momentum transport in stream self-intersection shocks, leading to larger debris pericenter distances and smaller apsidal precession angles; and large contrasts in the aspect ratios of the outgoing and incoming streams, which lead to a smaller fraction of the outgoing streams participating in self-intersection shocks:

\begin{itemize}

    \item \textit{Angular momentum transport}: As the stream self-crossing radius increases, collisions between streams at larger radii are not sufficiently violent to significantly perturb the incoming streams. As a result, angular momentum transfer becomes smoother, rather than episodic and violent as in the early-time strong-shock regime. The increasing pericenter distance of the incoming debris due to angular momentum transport is clearly shown in the right panel of Figure~\ref{fig:1t0_shock}. As gas falls inward for the first time, its pericenter $r_{\rm min} \propto j^2$ grows from $\sim 10$\,$r_g$ at a few $1000$\,$r_g$ to several tens of $r_g$ after passing through the self-intersection shock. Because a larger pericenter produces weaker apsidal precession, the system begins to resemble a weakly relativistic TDE \citep{2023ApJ...957...12R}.

    \item \textit{Aspect ratio contrasts}: Passage through the nozzle shock both heats the gas and broadens its angular momentum distribution. Consequently, in the post-shock flow it expands both vertically and laterally, creating large contrasts in aspect ratio and density between the incoming and outgoing streams. As shown in the left panel of Figure~\ref{fig:1t0_shock}, at $1$\,$t_0$, the vertical scale height of the outgoing stream exceeds that of the unshocked incoming stream by more than a factor of $\sim 10$. The pronounced contrasts in aspect ratio and density are further illustrated in Figure~\ref{fig:weak_stream_1t0}, which illustrates how the outgoing stream slips through the incoming stream. As a result of the vertical expansion, only a small fraction of the outgoing stream participates in the shock. 
    
\end{itemize}

\subsection{Orbital Evolution} \label{subsec:circular}
\begin{figure}[htb!]
    \centering
    \includegraphics[width=1.0\linewidth]{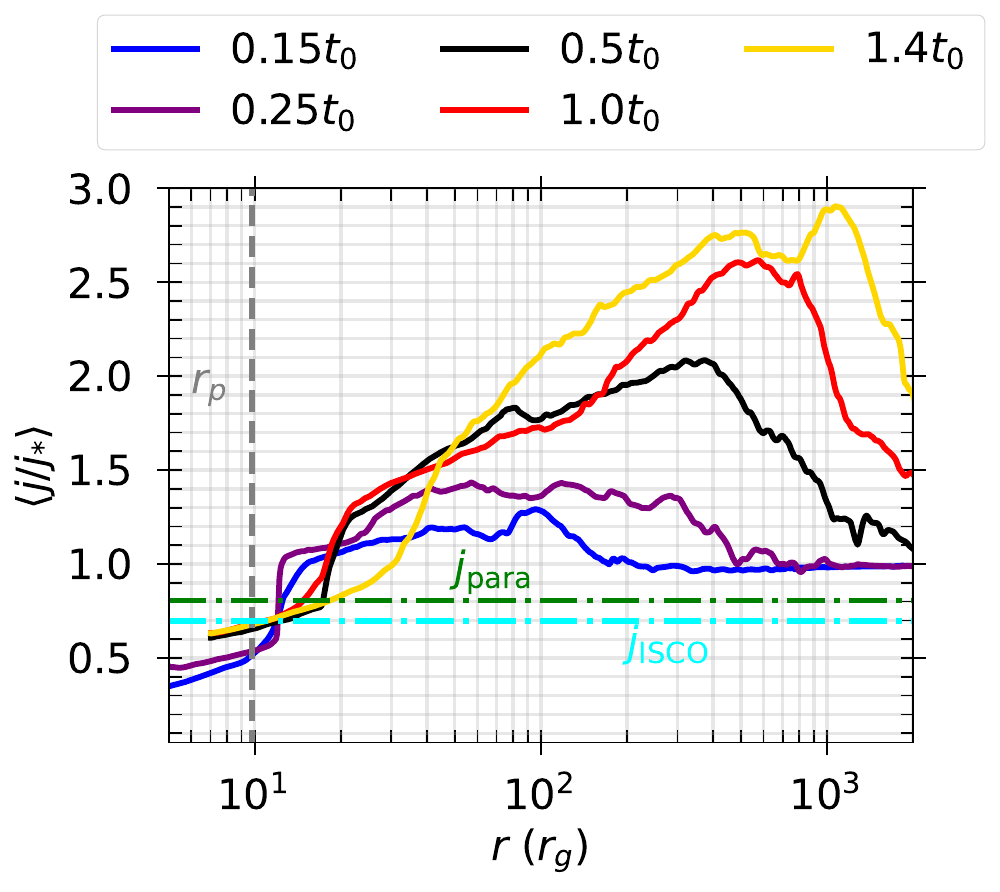} 
    \caption{Specific angular momentum relative to the initial stellar angular momentum profiles. All quantities are further mass-weighted. We also indicate the ratio of the ISCO angular momentum $(j_{\mathrm{ISCO}}$, cyan) and the angular momentum required for direct capture under a parabolic orbit $(j_{\mathrm{para}}$, green) for reference. \label{fig:tde_jang}}
\end{figure}
\begin{figure}[htb!]
    \centering
    \includegraphics[width=1.0\linewidth]{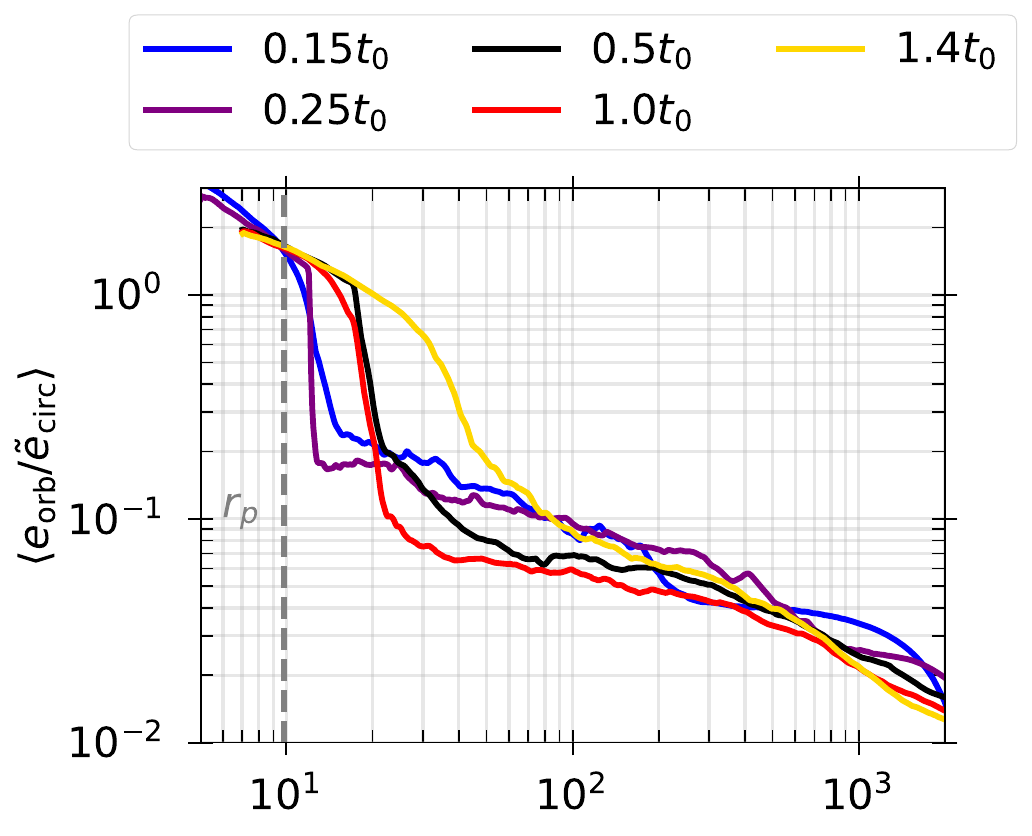} 
    \caption{Specific orbital energy as a function of distance from the SMBH, normalized by the energy required for circularization of debris into a circular disk with size of $2\,r_{p}$, which is $GM_{\rm BH}/[4r_{p}]$. All quantities are mass-weighted. \label{fig:tde_eorb}}
\end{figure}
\begin{figure}[htb!]
    \centering
    \includegraphics[width=1.0\linewidth]{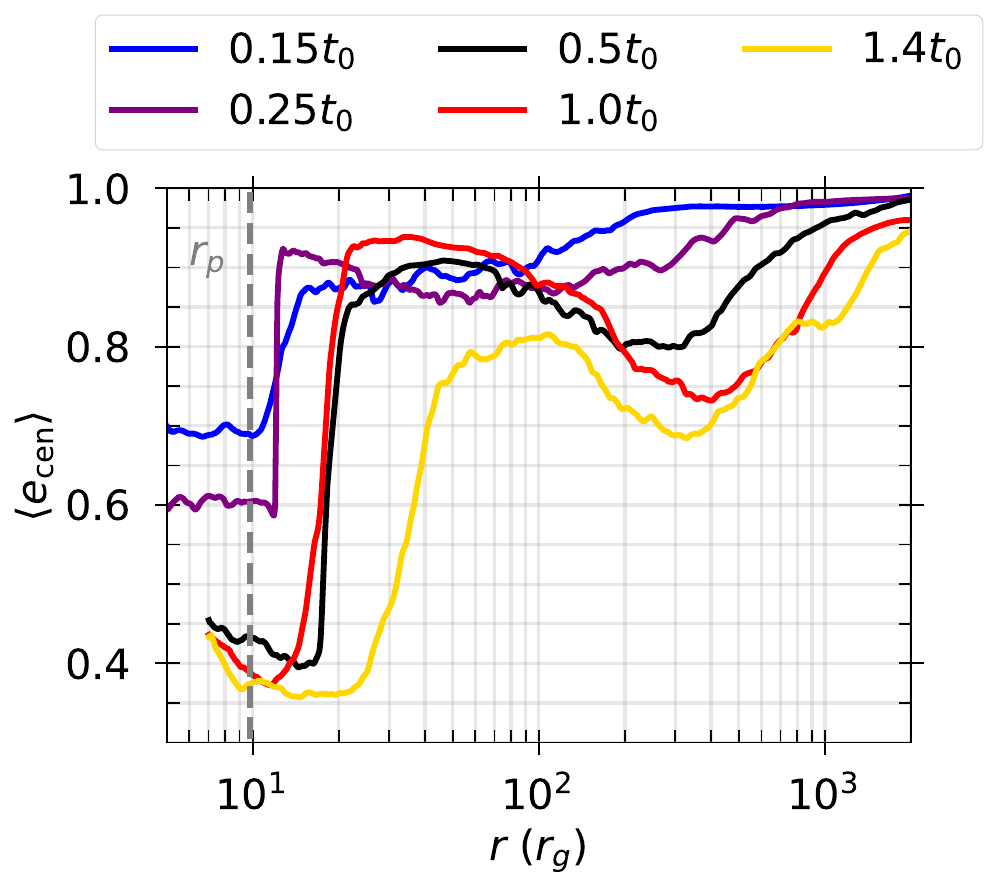} 
    \caption{Mass-weighted average of the orbital eccentricity of bound debris as a function of distance from the supermassive black hole. \label{fig:ecen}}
\end{figure}

In the preceding subsection, we discussed how shock-mediated changes in the debris' angular momentum and energy affect subsequent debris shocks. Here we change focus to consider how these changes affect the debris' orbits. Figure~\ref{fig:tde_jang} shows the density-weighted specific angular momentum $j$ of the debris, normalized by the initial stellar value $j_{*}$, as a function of distance from the SMBH at several different times. As the event proceeds, the system arranges itself so that fluid with greater $j$ occupies the outer regions. We also find that $j$ within the initial $r_{p}$ is generally lower than either the $j$ associated with a radial plunging orbit or the $j$ of the innermost stable orbit (ISCO). This fact implies that this fluid should soon accrete onto the SMBH. 

In Figure~\ref{fig:tde_eorb}, we present the ratio $e_{\rm orb}/\tilde{e}_{\rm circ}$ as a function of radius at the same times as in Figure~\ref{fig:tde_jang}. Here $e_{\rm orb}$ is the mass-weighted mean orbital energy and $\tilde{e}_{\rm circ} \equiv - GM_{\rm BH}/4r_p$ is the energy of an orbit with a semi-major axis $2\,r_p$, which is also the radius of a circular orbit whose angular momentum is the star's specific angular momentum.

Note that for a fixed angular momentum, eccentric orbits are less bound to the SMBH than circular orbits. Except for the small amount of matter located close to the SMBH (which has $e_{\rm orb}/\tilde{e}_{\rm circ} \gtrsim 1$), the rest of the debris is much more weakly bound than test particles on a circular orbit of radius $2\,r_p$. As demonstrated in Figure~\ref{fig:mrmdot}, even at $1.4\,t_0$, the fraction of the mass within $10^3\,r_g$ is $\lesssim 10^{-2}$\,$M_{\odot}$, whereas the binding energy of the gas at this radius or larger is $\lesssim 2 \times 10^{-2}\,\tilde{e}_{\rm circ}$.

As we have already shown, the distribution of debris angular momentum also changes during the event. The combined changes in the debris energy and angular momentum cause an evolution in the debris' orbital eccentricity. We define the specific eccentricity $e_{\rm cen}$ of a fluid element as
\begin{equation} 
\begin{aligned}
    e_{\rm cen} = \sqrt{1 + 2e_{\rm orb}j^{2}}.
\end{aligned}
\end{equation}
Its mass-weighted mean on spherical shells is shown in Figure~\ref{fig:ecen}. At $t\simeq0.15\,t_{0}$, when early-returning debris starts to be shocked, the eccentricity is generally high, with $e_{\mathrm{cen}}\approx 0.7-0.85$ near the SMBH and $\approx 0.95$ at larger radii. As more debris returns and experiences shocks, the eccentricity for the material near the initial $r_{p}$ decreases, reaching $\approx 0.4$ at $t \simeq 0.5\,t_{0}$; at later times, the eccentricity close to $r_p$ changes rather little. On the other hand, in the bulk of the debris, which remains much farther out, the eccentricity decreases throughout the event, but slowly, dropping only to $0.7-0.9$ at $t \simeq 1.4\,t_{0}$. The sharp jump in eccentricity at $r\simeq 10-20$\,$r_g$ for $t\gtrsim 0.25$\,$t_0$ in all the profiles presented above marks the approximate position of the orbital pericenter of the newly incoming stream after undergoing angular momentum transport. As this figure shows, the fresh debris pericenter distance expands from $\gtrsim 10\,r_g$ at early times to $\simeq 30\,r_g$ at $t = 1.4\,t_0$.

\subsection{Electromagnetic Observables} \label{subsec:emobserve}
\begin{figure}[htb!]
    \centering
    \includegraphics[width=1.0\linewidth]{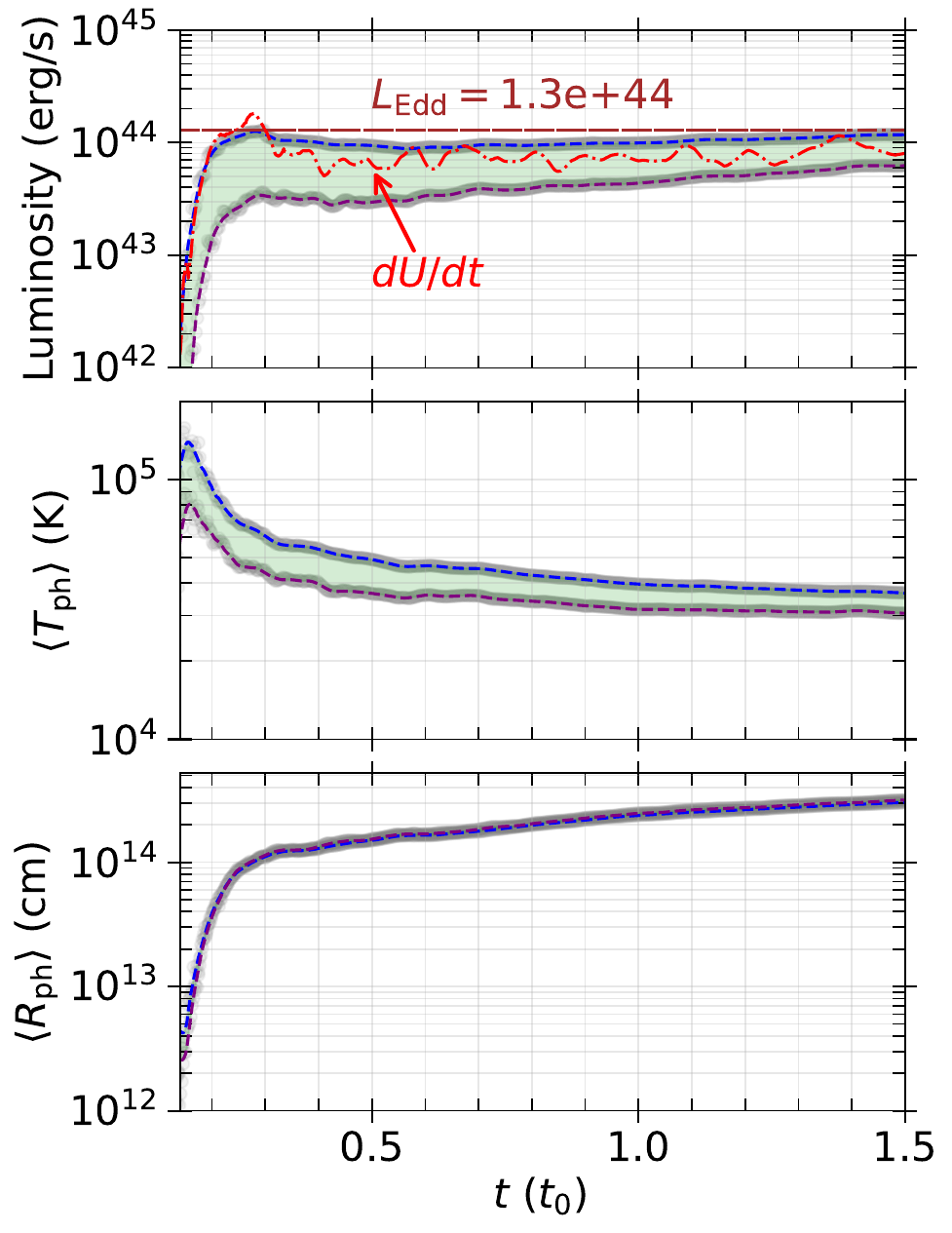} 
    \caption{
    Temporal evolution of the luminosity (top), the luminosity-weighted average photospheric temperature (middle, $\langle T_{\mathrm{ph}}\rangle$), and the average photospheric radius (bottom, $\langle R_{\mathrm{ph}}\rangle$). Short-dashed lines are moving averages of the raw data shown in gray.  In the top panel, a horizontal brown long-dashed line marks the Eddington luminosity, and the red line represents a moving average of the gross heating rate shown in Figure~\ref{fig:time_series_summary}. In all three plots, a short-dashed blue line indicates the results obtained including all contributions within the photosphere while a short-dashed purple line includes only contributions within the thermalization sphere; the region between them is shaded in green.  At $t = 0.3\,t_0$, the more inclusive luminosity estimate is larger by a factor $\simeq 3$, and this ratio decreases steadily with time. For $R_{\rm ph}$, the two calculations coincide. \label{fig:light_curve}}
\end{figure}

Although this work does not include explicit radiative transfer calculations, we can still roughly estimate the emergent luminosity, photospheric radius, and temperature by using the thermal energy content and the photon diffusion time. Here, we discuss the main findings and limitations of our estimates, while a detailed description of the procedure is provided in Appendix \ref{sec:estimate}.

As shown in Figure~\ref{fig:light_curve}, the total luminosity $L$ closely tracks the time derivative of thermal energy $dU/dt$ (red): $L$ rises rapidly to $10^{44}$\,erg\,s$^{-1}$ up to $0.3\,t_{0}$ ($\simeq 7$\,days) and plateaus at that level, which is close to the Eddington limit. The luminosity-weighted\footnote{To test the robustness of the temperature and photospheric radius estimates, we compared two averaging methods: a solid-angle-weighted average and a luminosity-weighted average, but their differences are very small.} blackbody temperature (see Appendix \ref{sec:estimate}) gradually decreases from its maximum of $\approx 1 \times 10^5$\,K at $t \simeq 0.15\,t_0$ to $40,000$\,K at $t \simeq 1.5\,t_0$, as illustrated in the middle panel of Figure~\ref{fig:light_curve}. Meanwhile, the photospheric radius (bottom panel), estimated using the Stefan-Boltzmann law $\langle R_{\rm ph}\rangle =\sqrt{L/[4\pi\sigma \langle T\rangle ^4]}$, rises rapidly until $t \lesssim 0.3\,t_0$, but then more slowly thereafter, reaching $2.5\times10^{14}$\,cm $\sim 1700\,r_g$ at $t=1.5\,t_{0}$. The overall behaviors of both the luminosity and temperature described above are consistent with the debris evolution in our simulation --- early strong shocks lead to a sharp change in the observables, while transition to a weaker and less variable shock regime yields more gradual changes.

We caution that the light curve shape and the debris dynamics we found could be significantly affected by the fact that we do not take into account radiation energy loss.\footnote{We note that in the radiation-hydrodynamic simulations of relativistic TDEs with similar parameters by \citet{2025arXiv251212985H}, the photon diffusion time near the circularization radius is $O(10)$ days prior to the peak mass-return time ($\simeq 15.5$\,days) and remains $O(10)$ days at $200-600\,r_{g}$ around the peak. This suggests that our assumption of thermal equilibrium between radiation and fluid may be justified.} In particular, the gross heating rate remains within a factor of a few of the estimated luminosity, suggesting a rough balance between radiative cooling and heating. If so, our assumption of adiabatic behavior is questionable. This same approximation may lead to overestimation of the radiation pressure. Another important systematic uncertainty lies in the estimation of the photon diffusion time. Our estimate relies on a globally estimated density scale height (Eqn.~\ref{eqn:rhosc}) and optical depth. Any inhomogeneities complicate the estimation of the photon diffusion time.

\section{Discussion} \label{sec:discuss} 

\subsection{Suppression of Relativistic Apsidal Precession}
Our main result is that the overall debris evolution in our simulation of a strongly relativistic TDE is qualitatively similar to that in \citet{2023ApJ...957...12R} for a weakly relativistic TDE: the debris remains highly-eccentric well past the peak mass-return time, and circular disk formation proceeds very slowly. In addition, the luminosity is primarily powered by shocks, rather than accretion.

This result is surprising because it has been widely expected that the strong apsidal precession associated with highly-relativistic events (i.e., those with $r_p \lesssim 10\,r_g$) should lead to rapid circularization on the order of the peak mass-return time. This in itself would release a very large amount of energy because the binding energy of a circular orbit at $\sim 10\,r_g$ is far greater than the original binding energy of the debris.

The central reason for the striking difference is that the conventional picture omitted the angular momentum exchange resulting from large-angle collisions between infalling and outgoing streams. In these shocks, the incoming stream gains angular momentum, enlarging its pericenter distance and thereby diminishing the relativistic apsidal precession and weakening subsequent shocks. In fact, $\Delta\phi_{\rm precess} \propto r_p^{-1} \propto j^{-2}$. In other words, for a given fractional angular momentum change (in the linear regime), the fractional change in the apsidal precession is twice as large.

We emphasize that the strong-shock phase does exist, but it is short-lived and lasts only $\simeq 0.3\,t_0$, about a week for our simulation's parameters. For the rest of the event, shocks overall weaken, drastically slowing the entry of debris mass into a circular compact disk.
 
\subsection{Unification of Relativistic and Non-relativistic Events}
One observational implication of the resemblance between weakly and strongly relativistic TDEs is that a single physical model \citep{Krolik+2025,piran2026tdepopulationfirstprinciplesmodels} can describe how the TDE peak luminosity depends on $M_*$ and $M_{\mathrm{BH}}$, almost regardless of how strong relativistic effects are. There are only two exceptions. One is the so-called Extreme TDEs \citep{2023ApJ...946L..33R} that take place in the extremely relativistic regime, where the orbital pericenter is so close to the black hole ($r_p\sim 6\,r_g$) that the apsidal precession can exceed $180^{\circ }$. In this limit, rather than the irregular eccentric shape of most TDEs, the debris morphology becomes roughly axisymmetric. Such extreme events should be rare for $M_{\rm BH}\lesssim 10^{7}\,M_{\odot}$, but are the majority of the TDE population when $M_{\rm BH}\gtrsim$ a few $10^{7}\,M_{\odot}$, and are nearly the entirety of the events (all but those involving the highest-mass stars) when $M_{\rm BH} > 1 \times 10^8\,M_\odot$ \citep{piran2026tdepopulationfirstprinciplesmodels}. The other is the possible effect of Lense-Thirring torques when the black hole spins rapidly, the star's orbit is oblique to the spin, and the pericenter is small enough to make Kerr effects strong \citep[see, e.g., the preliminary explorations by][]{2015ApJ...809..166G, 2019MNRAS.487.4790G,2024ApJ...971L..46P}.

\subsection{Slow `Circularization' Rate} \label{subsec:circularization}
As we have previously remarked, `circularization' as it is used in the TDE literature refers to two effects of the same process, namely the dissipation of orbital energy into heat. As the orbital binding energy grows, the debris' semi-major axis shrinks and, for fixed specific angular momentum, so does its orbital eccentricity.

To quantify the circularization rate, we follow \citet{2023ApJ...957...12R} by defining its efficiency as
\begin{equation} \label{eqn:eff}
    \eta = \frac{dU/dt}{M_{\rm shocked}\tilde{e}_{\rm circ}/t_{0}}.
\end{equation}
Here, $dU/dt$ is the gross heating rate of the fluid (yellow line in the bottom panel of Figure~\ref{fig:time_series_summary}) and $M_{\mathrm{shocked}}$ denotes the mass of the shocked gas.\footnote{To identify shocked fluid, we utilize the pseudo-entropy log${\mathcal{K}}$ such that $\mathcal{K} =  (P/\rho ^{\Gamma _{\mathrm{eff}}})$, where $\Gamma_{\mathrm{eff}}$ is the effective adiabatic index accounting for both the fluid and radiation pressure. We find that a cutoff value of $1.15$ effectively separates the low-entropy (unshocked) fluid from the high-entropy (shocked) fluid. This cutoff also yields a progressive accumulation of shocked mass over time, closely tracking the cumulative distribution of the fallback rate. For further details, see Appendix \ref{sec:distribution}.} The time evolution of $\eta$ has already been shown in Figure~\ref{fig:one_figure_summary}. This definition characterizes the efficiency with which the debris orbit’s semi-major axis decreases from $\sim (M_{\mathrm{BH}}/M_*)^{1/3}r_p$ toward $\sim r_p$.

We find $\eta$ can be as large as $\sim 3$ very early in the event ($t\lesssim 0.3\,t_{0}$). In other words, the very small amount of mass that returns during this early stage can be immediately circularized. However, when $t\gtrsim 0.3\,t_{0}$, $\eta$ decreases rapidly, reaching $\approx 0.2$ by $t\simeq 0.4\,t_0$, and flattening out at $5 - 10\,\%$ after $t \simeq 0.7\,t_0$. Thus, the  circularization pace is $10 - 20$ times slower than what is conventionally expected---complete circularization within $1\,t_{0}$. This sharp slowdown in `circularization' occurs because the strong shocks that occur when the apsidal precession angle is large dramatically weaken when this angle is diminished as a result of the incoming debris acquiring additional angular momentum before reaching an enlarged orbital pericenter.

\subsection{Comparison to Previous Studies of Strongly Relativistic TDEs}

Only a handful of numerical studies have simulated strongly relativistic TDEs, either starting from the star’s initial approach or focusing on the returning debris whose properties are assumed for a parabolic trajectory.

\citet{2022MNRAS.510.1627A} investigated the debris evolution and disk formation in a strongly relativistic TDE with $r_p=7\,r_g$ over a duration of $5 - 7$\,days (corresponding to $0.21 - 0.30$\,$t_0$), overlapping with the early phase of our simulation. They likewise observed substantial disruption of the incoming stream caused by strong self‑intersection shocks (their Figure $3$). They also found that the debris remains highly eccentric ($e_{\mathrm{cen}}\simeq 0.88$) until the end of their simulation; their radially averaged eccentricity profiles at $2.6$ and $6$\,days (their Figure $17$) are strikingly similar to our profile at $0.15\,t_0\simeq 3.5$\,days shown in Figure~\ref{fig:ecen}. Because of the short evolution duration, their simulation may not capture the complete transition from the strong‑shock phase to the weaker, saturated‑shock phase, which takes place a week after the disruption. However, the minimal evolution of their radial eccentricity profiles between $2.6$ and $6$\,days indicates that the debris evolution might have already entered the saturation phase. In short, although their simulations cover only a brief period, they align closely with ours during the timeframe in which the two overlap. 

\citet{2024ApJ...971L..46P} performed long-term simulations of both weakly and strongly relativistic TDEs. Their relativistic simulation adopted $r_{p}=10\,r_{g}$ and followed the evolution for one year. Because the description of the debris evolution in their relativistic case is quite brief, and because their parameter space (tilted TDE with a spinning SMBH) is rather different than ours, a direct comparison with our results is not straightforward. Nonetheless, we want to point out that their luminosities, blackbody temperatures, and blackbody radii, estimated by solving the radiative transfer equation in post-processing, appear to be very similar to our predictions, despite the caveats of our estimates.

Most recently, using a radiation‑hydrodynamic simulation adopting a stream‑injection scheme and a pseudo-Newtonian potential that captures relativistic apsidal precession, \citet{2025arXiv251212985H} studied a strongly relativistic event ($r_p \simeq 13\,r_g$) up to $\simeq 2.3$\,$t_0\simeq 35$\,days since disruption (their $t_{0}\simeq 15.5$\,days). Similar to our results, they found that shocks power the event and circularization is sufficiently slow for debris to remain eccentric past the peak mass-return time (their Figure $4$). However, circularization appears to proceed faster in their simulation than in ours. Near the black hole, a quasi-circular accretion flow with mass $\simeq 10^{-2} - 10^{-1}\,M_{\odot}$\footnote{The mass is estimated assuming a mid-plane density of $10^{-9}-10^{-10}$\,g\,cm$^{-3}$ (their Figure $1$) and an aspect ratio of $0.4-0.5$ (their Figure $7$) at $\simeq 25$\,days} and $e_{\rm cen} \simeq 0.2-0.4$ forms around $20 - 25$\,days ($1.3 - 1.6$\,$t_0$). But it is so hot that it might better be described as `quasi-spherical', and it extends out to $200\,r_{g}\simeq 15\,r_{p}$, an order of magnitude larger than the usual radius ($2\,r_{p}$) associated with `circularization'. In contrast, the gas within the same region in our simulation has mass of $10^{-3}\,M_{\odot}$ and eccentricities of $0.4-0.8$ at $1.4\,t_{0}$.


\section{Conclusion}\label{sec:conclusion}

In this work, we present a general relativistic 3D hydrodynamic simulation of a strongly relativistic TDE involving a realistic Sun-like star disrupted by a $10^{6}$\,$M_{\odot}$ non-spinning SMBH with a pericenter distance of $10$\,$r_{g}$ ($\beta = r_t/r_p=5$). The simulation self-consistently follows the debris evolution beyond the peak mass-return time, starting from the star's initially parabolic approach. We focus on studying the debris morphology and shock formation, and estimating the electromagnetic observables.

Our main results can be summarized as follows:
\begin{itemize}
        
    \item \textit{Overall weak impact of relativistic effects}: In the conventional picture, strong relativistic effects lead to prompt disk formation in relativistic TDEs. Our simulation instead finds that although relativistic effects initially generate powerful shocks that dissipate kinetic energy efficiently, this phase lasts only about a week ($\sim 0.3$\,$t_0$). After this time, the dissipation rate drops to a lower level as the pericenter distances of the newly incoming streams increase due to angular-momentum transfer mediated by continuous self-intersection shocks. Consequently, the overall debris evolution near the peak mass-fallback time becomes qualitatively similar to that in a weakly relativistic TDE.
    
    \item \textit{Slow circularization}: The returned debris remains eccentric past the peak mass-return time, and most of the mass settles near the characteristic apocenter distance ($\simeq10^{3}\,r_{g}$), well beyond the circularization radius ($2\,r_{p}$). We estimate that circularization completes only by $10 - 20$\,$t_0 \simeq 230 - 470$\,days after the disruption, substantially longer than the expected circularization on the mass-return timescale of $1\,t_{0}\simeq 20 - 30$\,days.
    
    \item \textit{Shocks as the main power source}: Nozzle and self intersection shocks, not accretion, primarily power the event. We find that the overall ranges of the bolometric luminosity ($\lesssim 10^{44}$\,erg\,s$^{-1}$) and blackbody temperatures ($2-4\times 10^{4}\,K$) are consistent with observations, closely tracking the hydrodynamic evolution and thermal energy dissipation. However, we also find that radiative cooling becomes more important in the later phase of the evolution, emphasizing the importance of a self-consistent treatment of radiation transfer.

\end{itemize}

Thus, this work suggests that there is a fundamental unity of mechanism linking all TDEs of main-sequence stars, excepting only those extreme TDEs with such small pericenter distances that the star is forced to loop around the black hole several times and possibly small pericenter events involving rapidly spinning black holes whose axes are oblique to the stellar orbital axis. This unity supports the construction of a unified TDE population model \citep{piran2026tdepopulationfirstprinciplesmodels}.

\begin{acknowledgments}
Ho-Sang (Leon) Chan acknowledges support from the Croucher Scholarship for Doctoral Studies by the Croucher Foundation. T. Ryu is grateful to Sasha Tchekhovskoy and Pavani Jairam for helpful discussions and comments. This work used Stampede3 at the Texas Advanced Computing Center (TACC) through allocation PHY240321 (PI: T. Ryu) from the Advanced Cyberinfrastructure Coordination Ecosystem: Services \& Support (ACCESS) program, which is supported by U.S. National Science Foundation grants \#2138259, \#138286, \#2138307, \#2137603, and \#2138296. This work was partially supported by: an ERC advance grant MultiJets and the Simon SCEECS collaboration (TP); and NASA TCAN grant 80NSSC24K0100 (JHK).
\end{acknowledgments}

\appendix

\section{Estimating the Electromagnetic Observables} \label{sec:estimate} 
\begin{figure}[htb!]
    \centering
    \includegraphics[width=1.0\linewidth]{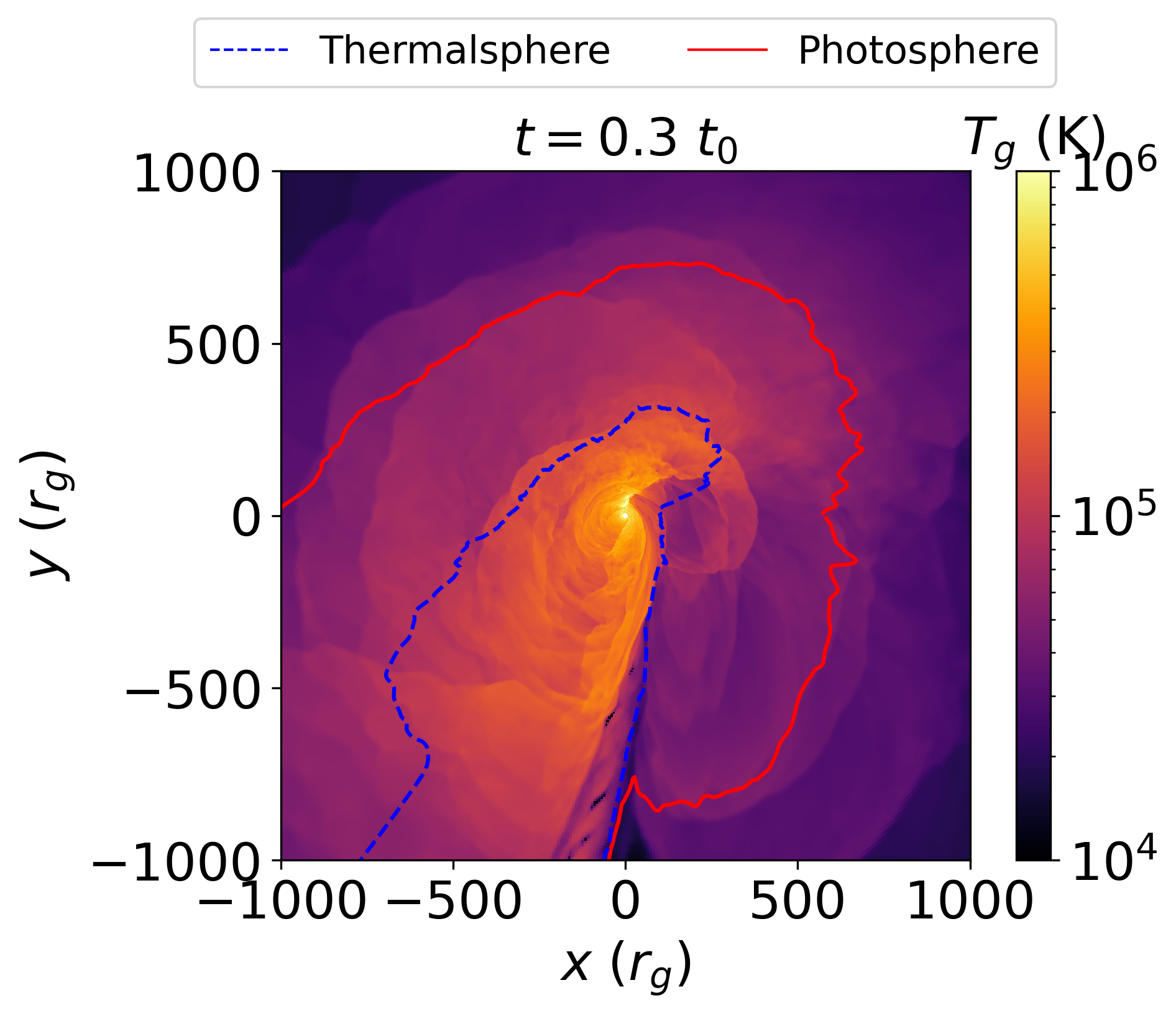} 
    \caption{Temperature map at $0.3\,t_0$ (when the luminosity peaks and subsequently saturates) in the $x-y$ plane. The photosphere and thermalization surface are shown with red solid and blue-dashed curves, respectively. \label{fig:thermalization}}
\end{figure}
\begin{figure}[htb!]
    \centering
    \includegraphics[width=1.0\linewidth]{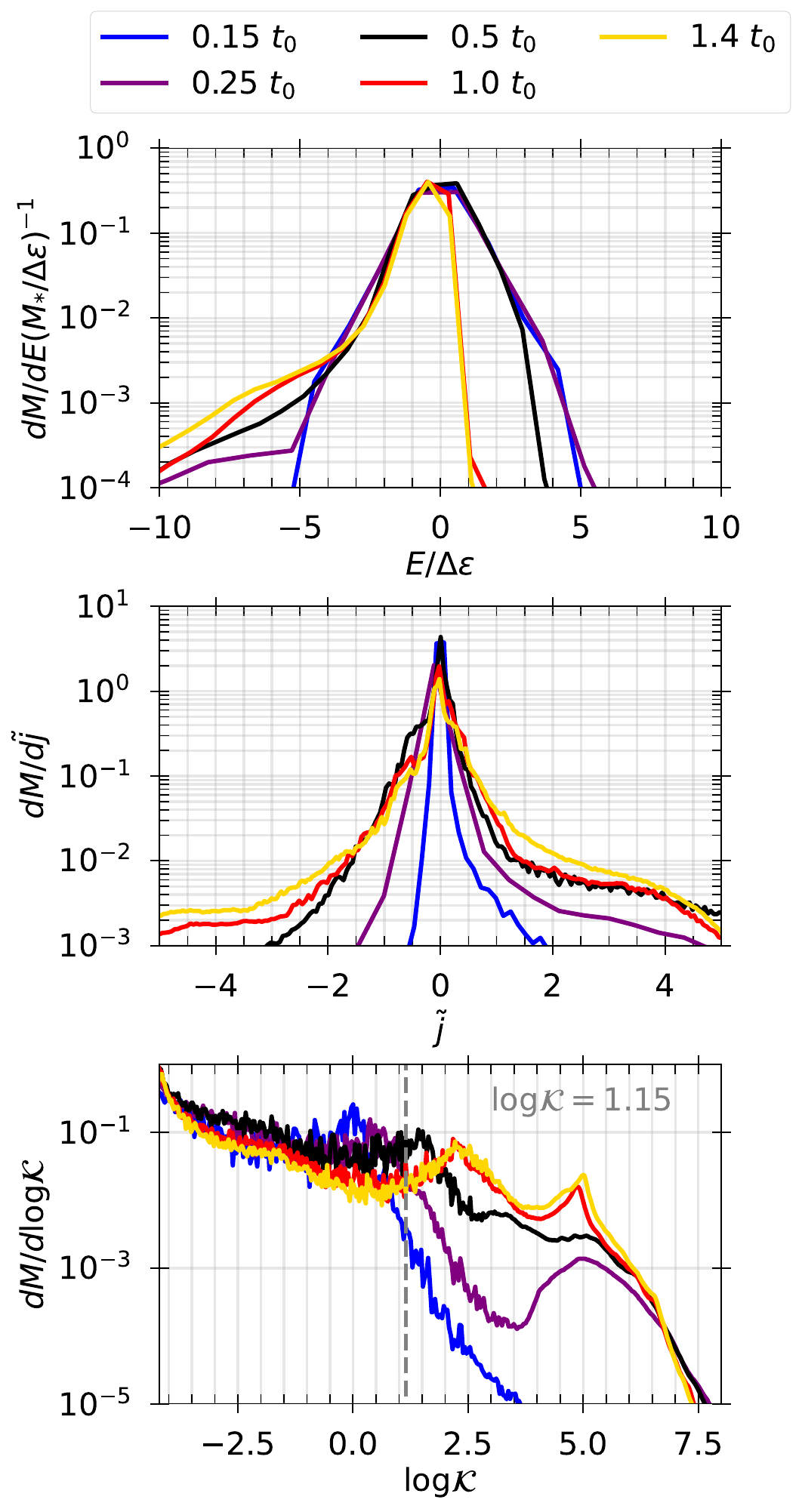} 
    \caption{Orbital energy (top), relative specific angular momentum with respect to the star’s initial angular momentum (middle), and the pseudo-entropy (bottom) at different epochs. In the bottom row, the vertical gray line marks log$\mathcal{K} = 1.15$. \label{fig:distribution}}
\end{figure}
\begin{figure}[htb!]
    \centering
    \includegraphics[width=1.0\linewidth]{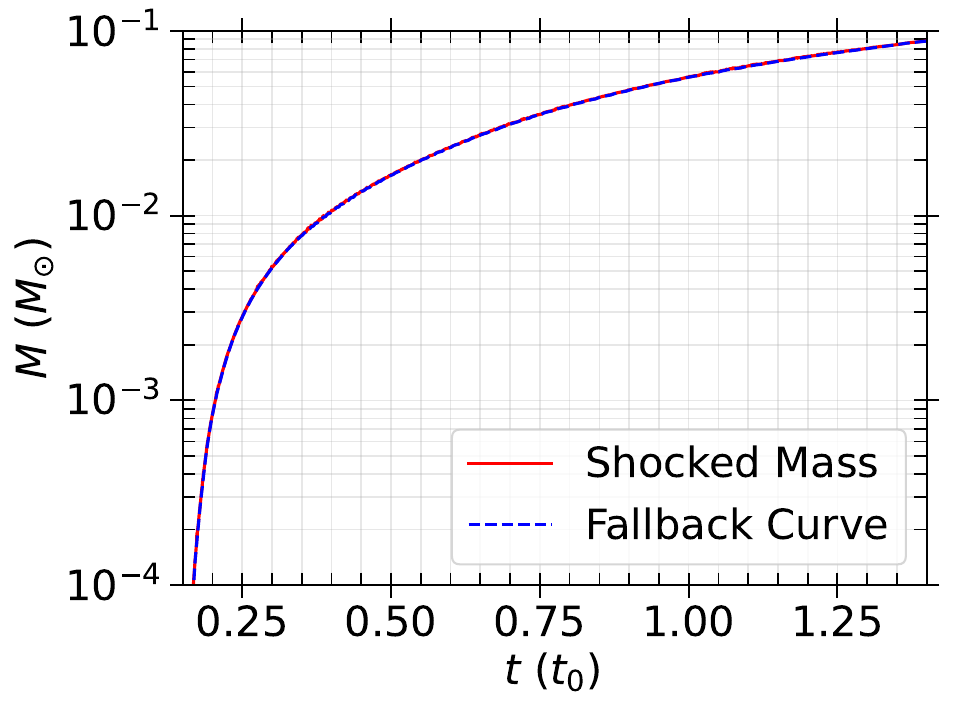} 
    \caption{Total gas mass with log$\mathcal{K} > \text{log}\mathcal{K}_{\mathrm{cut}} = 1.15$ as a function of time (red) The fallback curve (blue dashed) is identical. \label{fig:shockmasstime2}}
\end{figure}
\begin{figure*}[htb!]
    \centering
    \includegraphics[width=0.45\linewidth]{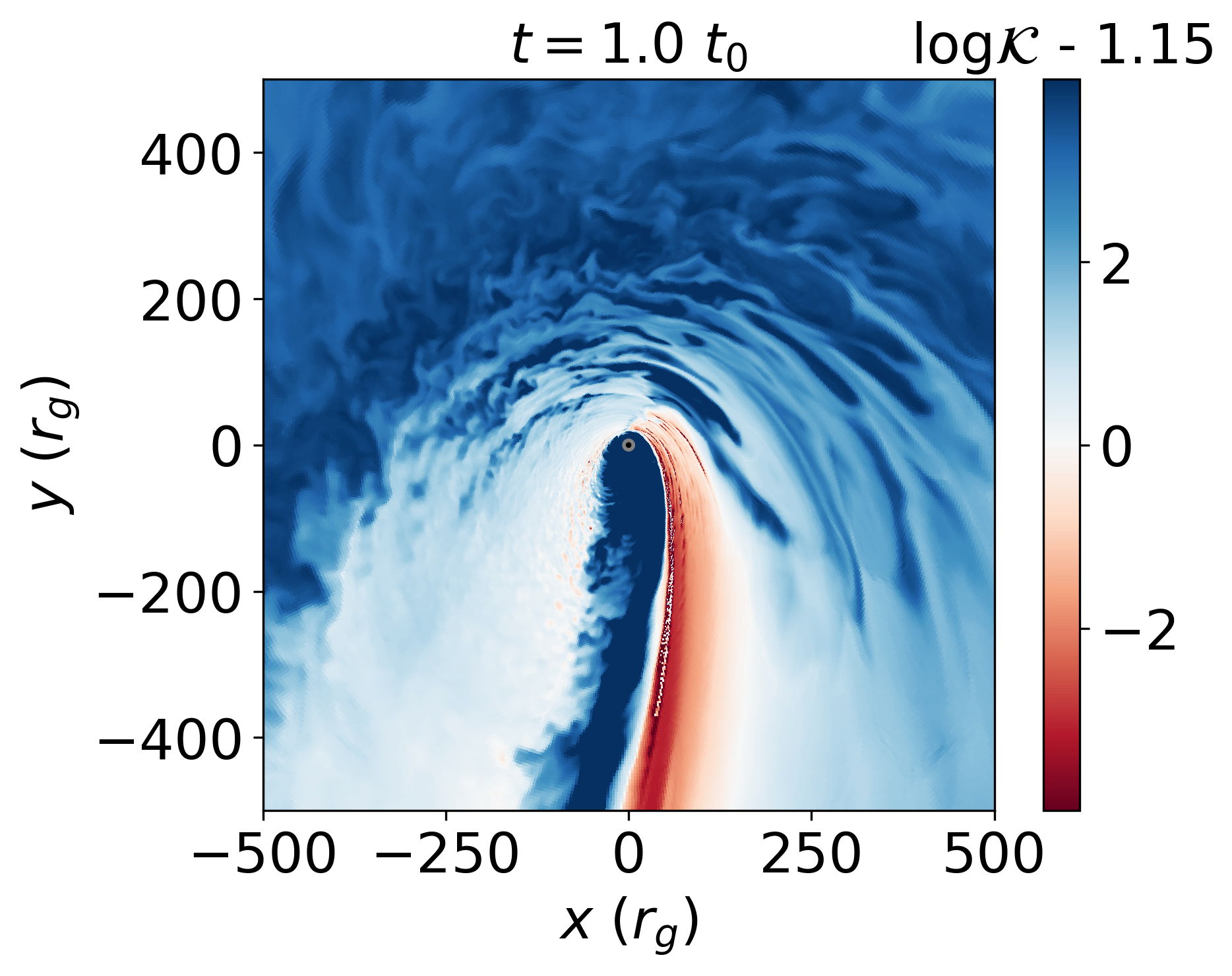}
    \includegraphics[width=0.45\linewidth]{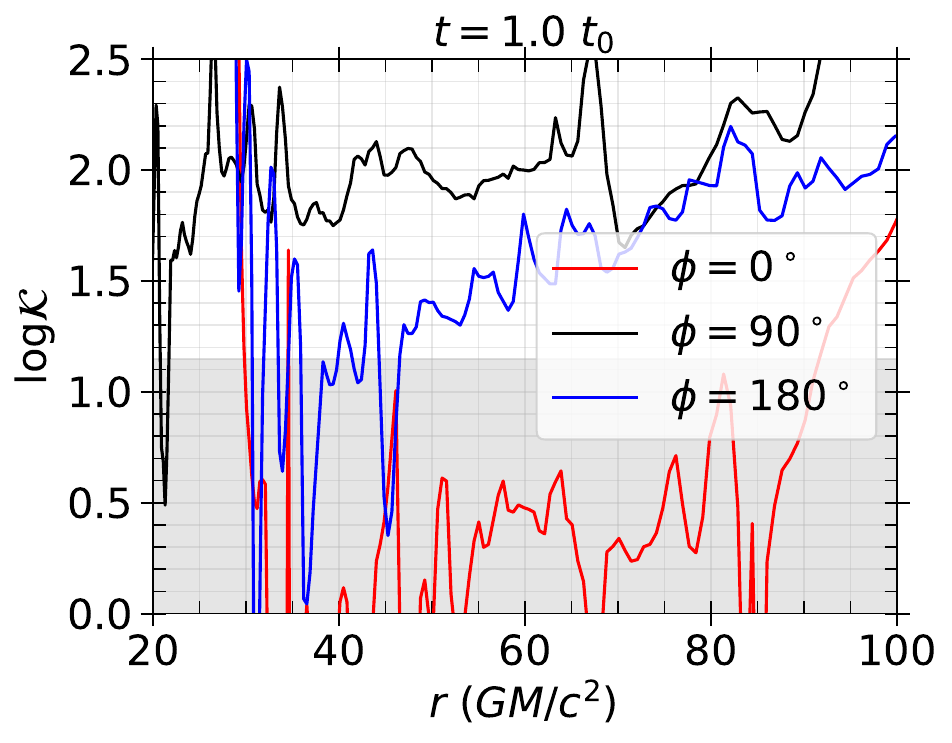}
    \caption{(Left) Map of log$\mathcal{K}_{\mathrm{cut}} - 1.15$ in the $x-y$ plane at $1\,t_0$. (Right) Same as the left panel, but measuring the radial variations of log$\mathcal{K}$ along different azimuthal angle $\phi$. The grey shaded region represent log$\mathcal{K}_{\mathrm{cut}} < 1.15$. \label{fig:shockmasstime}}
\end{figure*}

To estimate the electromagnetic observables, for each snapshot, we first identify the photospheric radius, $R_{\mathrm{ph}}$, defined as
\begin{equation} 
    \tau(R_{\rm ph}) = \int_{R_{\rm ph}}^{R_{\rm out}} \rho (\kappa_{\rm sc} + \kappa_{\rm abs})dr = 1,
\end{equation}
where $R_{\mathrm{out}}$ denotes the outer radial boundary, and $\kappa _{\mathrm{sc}}=0.34$\,$\mathrm{cm^{2}}$\,$\mathrm{g^{-1}}$ is the Thomson scattering opacity, assuming a hydrogen mass fraction of $0.7$. The absorption opacity, $\kappa_{\mathrm{abs}}$, is computed (interpolated) using the OPAL opacity tables for solar metallicity \footnote{In cases where the target cells fall outside the tabulated range, we instead adopt the analytical expression for the absorption opacity given by \citet{2017MNRAS.471.3200M, 2024ApJ...971L..46P}.}. Although we account for both scattering and absorption opacity, electron scattering dominates over absorption processes, resulting in the scattering photosphere becoming equal to the total absorption photosphere.

We then estimate the cooling time of a fluid element located at radius $r$ within the photosphere as
\begin{equation} 
    t_{\rm cool}(r) = \frac{h_\rho \tau(r)}{c},
\end{equation}
where $h_\rho$ is the density-weighted scale height along radial paths
\begin{equation} \label{eqn:rhosc}
    h_\rho = \int_{r}^{R_{\rm ph}}\rho s ds/\int_{r}^{R_{\rm ph}}\rho ds.
\end{equation}
The flux at the photosphere along a radial path is estimated as
\begin{equation} \label{eqn:luminosities}
    \frac{dL}{dA} = \int_{R_{\rm diff}}^{R_{\rm ph}} \frac{aT^{4}}{t_{\rm cool}}\frac{s^{2}}{R_{\rm ph}^{2}}ds,
\end{equation}
where $R_{\rm diff}$ is the radius at which the cooling time equals the system evolution time. Setting the lower limit of the integral as $R_{\rm diff}$ ensures that the radiation that can escape within the simulation time can contribute to the total luminosity. If $R_{\rm diff} > R_{\rm ph}$, we set the differential luminosity to zero. Additionally, if $R_{\mathrm{ph}} \geq 0.8$\, $R_{\mathrm{out}}$, we set the differential luminosity to zero in order to avoid contamination from the finite radial domain. This ensures that our luminosity estimates are not artificially inflated by boundary effects.

We also performed a more conservative estimate in which we excluded contributions from above the thermalization sphere, i.e., thereby changing the upper limit on the integral in Equation~\ref{eqn:luminosities} from $R_{\rm ph}$ to $R_{\rm th}$. The thermalization radius $R_{\rm th}$ is computed from the relation
\begin{equation} 
    \tau(R_{\rm th}) = \int_{R_{\rm th}}^{R_{\rm out}} \rho \sqrt{\kappa_{\rm abs}(\kappa_{\rm sc} + \kappa_{\rm abs})} dr = 1.
\end{equation}
In Figure~\ref{fig:thermalization}, we compare the sizes of the photosphere and the thermalization surface at $t=0.3\,t_0$, which corresponds to the time when the luminosity peaks and subsequently saturates.

The total luminosity is then computed by integrating the differential luminosity along the radial direction
\begin{equation} 
    L = \int\int\frac{dL}{dA}\sin\theta d\theta d\phi,
\end{equation}
and the effective temperature of the photosphere is
\begin{equation} 
    T_{\rm eff} = \left( \frac{dL}{dA}\frac{1}{\sigma} \right)^{1/4},
\end{equation}
where $\sigma$ denotes the Stefan-Boltzmann constant. The differential luminosity-weighted average photospheric temperature is given by
\begin{equation} 
    \langle T\rangle = \int T_{\rm eff} \frac{dL}{dA}dA / \int \frac{dL}{dA}dA,
\end{equation}
while the solid angle-weighted average photospheric temperature is given by
\begin{equation} 
    \langle T\rangle = \int T_{\rm eff} d\Omega / \int d\Omega,
\end{equation}
where the integrals are performed over the photosphere.

Note that we neglect the effects of spacetime curvature near the SMBH.

\section{Energy, Angular Momentum, and Entropy Distribution} \label{sec:distribution} 

We show the orbital energy, specific angular momentum, and pseudo-entropy distributions in Figure~\ref{fig:distribution}. The orbital energy distribution is initially symmetric about $E = 0$. At later times, however, a larger fraction of the mass acquires more negative orbital energy, a consequence of shocks that convert orbital energy into thermal energy. The amount of mass with positive energy decreases over time simply because it corresponds to the initially unbound debris that escapes the computational domain.

The angular momentum distribution is initially narrowly peaked at and roughly symmetric about $j=j_*$, but it broadens through shock interactions, as we discussed in Section \ref{subsec:shock}. The increased amount of mass with higher angular momentum corresponds to the incoming stream, which attains larger orbital pericenters. In contrast, the material with lower angular momentum belongs to the outgoing stream, having lost angular momentum through stream self‑intersection.

For pseudo-entropy defined as $\mathcal{K}=(P/\rho^{\Gamma_{\rm eff}})$, the cut-off at log$\mathcal{K} = 1.15$ roughly separates shocked from unshocked material. The mass of high‑entropy material above this threshold rises sharply up to $\simeq 0.5t_0$ and more slowly at later times, a pattern consistent with the overall hydrodynamics of our simulation: namely, an initial phase of efficient circularization transitioning into saturated circularization.

In the left panel of Figure~\ref{fig:shockmasstime} we show the map of log$\mathcal{K} - 1.15$ in the $x-y$ plane at $t = t_0$, demonstrating that the cutoff value distinguishes the unshocked gas (deep red) from the shocked material (white/blue). In the right panel, we measure log$\mathcal{K}$ along the radial direction at the midplane for different $\phi$. The chosen cutoff value largely captures the outgoing shocked stream; the shaded region does not overlap with the blue curve for $r \gtrsim50\,r_g$. In Figure~\ref{fig:shockmasstime2}, we show that the log$\mathcal{K}$ cutoff value reasonably tracks the shocked mass, which approximately follows the fallback curve.


\bibliography{sample701}{}
\bibliographystyle{aasjournalv7}



\end{document}